\documentclass[journal]{IEEEtran}
\usepackage{graphicx,color}
\usepackage{cite}
\usepackage{setspace} 
\usepackage{amsmath}
\usepackage{amsmath,amsthm,amssymb}
\usepackage{multirow}
\usepackage{hhline}
\usepackage{epsfig}
\usepackage{subfigure}
\usepackage{epstopdf}
\usepackage{verbatim}
\usepackage{algorithm}
\usepackage{algorithmicx}
\usepackage{algpseudocode}
\usepackage{etoolbox}
\usepackage{cases}
\usepackage{csquotes}
\usepackage{enumitem}
\usepackage{mathtools,stmaryrd}
\SetSymbolFont{stmry}{bold}{U}{stmry}{m}{n}
\usepackage{bbm}
\usepackage{lettrine}
\usepackage{nicematrix}

\newcommand{\suchthat}{\;\ifnum\currentgrouptype=16 \middle\fi|\;}

\makeatletter
\newcommand*{\indep}{%
  \mathbin{%
    \mathpalette{\@indep}{}%
  }%
}
\newcommand*{\nindep}{%
  \mathbin{
    \mathpalette{\@indep}{\not}
  }%
}
\newcommand*{\@indep}[2]{%
  \sbox0{$#1\perp\m@th$}
  \sbox2{$#1=$}
  \sbox4{$#1\vcenter{}$}
  \rlap{\copy0}
  \dimen@=\dimexpr\ht2-\ht4-.2pt\relax
  \kern\dimen@
  {#2}%
  \kern\dimen@
  \copy0 
} 
\makeatother

\makeatletter
\newcommand*{\algrule}[1][\algorithmicindent]{%
  \makebox[#1][l]{%
    \hspace*{.2em}
    \vrule height .75\baselineskip depth .25\baselineskip
  }
}

\newcount\ALG@printindent@tempcnta
\def\ALG@printindent{%
    \ifnum \theALG@nested>0
    \ifx\ALG@text\ALG@x@notext
    \else
    \unskip
    \ALG@printindent@tempcnta=1
    \loop
    \algrule[\csname ALG@ind@\the\ALG@printindent@tempcnta\endcsname]%
    \advance \ALG@printindent@tempcnta 1
    \ifnum \ALG@printindent@tempcnta<\numexpr\theALG@nested+1\relax
    \repeat
    \fi
    \fi
}
\patchcmd{\ALG@doentity}{\noindent\hskip\ALG@tlm}{\ALG@printindent}{}{\errmessage{failed to patch}}
\patchcmd{\ALG@doentity}{\item[]\nointerlineskip}{}{}{} 
\makeatother

\newtheorem{theorem}{Theorem}
\newtheorem{prop}{Proposition}
\newtheorem{corr}{Corollary}

\allowdisplaybreaks

\begin{document}

\title{Chaotic Waveform-based Signal Design\\ for Noncoherent SWIPT Receivers}

\author{Authors}
\author{Priyadarshi Mukherjee, \textit{Member, IEEE}, Constantinos Psomas, \textit{Senior Member, IEEE}, and Ioannis Krikidis, \textit{Fellow, IEEE}
\thanks{P. Mukherjee is with the Advanced Computing \& Microelectronics Unit,  Indian Statistical Institute, Kolkata, India. C. Psomas and I. Krikidis are with the Department of Electrical and Computer Engineering, University of Cyprus, Nicosia 1678 (E-mail: priyadarshi@ieee.org, psomas@ucy.ac.cy, krikidis@ucy.ac.cy).

This work received funding from the European Research Council (ERC) under the European Union's Horizon 2020 research and innovation programme (Grant agreement No. 819819). It was also funded by the European Union's Horizon Europe programme (ERC, WAVE, Grant agreement No. 101112697),
and from the European Union HORIZON programme under iSEE-6G GA No. 101139291.}}

\maketitle
\begin{abstract}
This paper proposes a chaotic waveform-based
multi-antenna receiver design for simultaneous wireless information and power transfer (SWIPT). Particularly, we present a differential chaos shift keying (DCSK)-based SWIPT multiantenna receiver architecture, where each antenna switches between information transfer (IT) and energy harvesting (EH) modes depending on the receiver's requirements. We take into account a generalized frequency-selective Nakagami-m fading model as well as the nonlinearities of the EH process to derive closed-form analytical expressions for the associated bit error rate (BER) and the harvested direct current (DC), respectively. We show that, both depend on the parameters of the transmitted waveform and the number of receiver antennas being utilized in the IT and EH mode. We investigate a trade-off in terms of the BER and energy transfer by introducing a novel achievable `success rate - harvested energy' region. Moreover, we demonstrate that energy and information transfer are two conflicting tasks and hence, a single waveform cannot
be simultaneously optimal for both IT and EH. Accordingly, we propose appropriate transmit waveform designs based on the application specific requirements of acceptable BER or harvested DC or both. Numerical results demonstrate the importance of chaotic waveform-based signal design and its impact on the proposed receiver architecture.
\end{abstract}


\IEEEpeerreviewmaketitle

\section{Introduction}

With the advent of emerging technologies like massive machine-type communication and Internet of Things in recent years, the wireless traffic has been growing at a tremendous rate. Specifically, the growth is expected to be more than five times between $2019$ and $2028$ \cite{ericsson} with the data intensive applications witnessing approximately $1000$ times growth. Hence, in such times of ever increasing data traffic,  the overall network lifetime gets significantly affected due to limited battery constraints, especially in scenarios, where a large number of devices are deployed over a geographical region. Thereby, charging or powering these devices becomes a costly and critical concern. As a result, self-sustainable and low-powered next generation wireless communication networks are gaining importance as well as relevance in both academia and industry. In this context, the fact that radio frequency (RF) signals can also convey energy apart from information, the concept of wireless power transfer (WPT) and, in particular, of simultaneous wireless information and power transfer (SWIPT) is considered as a very promising and enabling technology \cite{swiptproc}.

The key idea of SWIPT is to extract both information and energy from the received RF signal. This is achieved by employing a rectifying antenna (rectenna) at the receiver, which converts the received RF signals to direct current (DC). Unlike conventional energy sources, where the available power for harvesting, in itself, is erratic in nature \cite{wptsurvey}, energy harvesting (EH) with SWIPT is a dedicated, controllable, continuous, and on-demand process. This joint extraction of information and energy is done by separating the information decoding and EH operations in space, in time, or in power \cite{procbruno}. The work in \cite{swiptmimo} explores SWIPT systems for multiple-input multiple-output broadcasting channel, where both separated and co-located EH and information decoding (ID) receivers are considered. The authors in \cite{swipt2} investigate the capacities of SWIPT systems with separate ID and multiple EH receivers. In this context, the aspect of accurate mathematical modelling of the EH circuit at the receiver plays a very important role. Some works propose simplified linear \cite{sir}, piece-wise linear \cite{rev1,plinear} and tractable logistic nonlinear model \cite{satm} of the EH circuit that originates from the saturation of the output power beyond a certain RF input power due to diode breakdown. The logistic model is obtained by fitting measurements from practical RF-based EH circuits for a given excitation signal and is certainly an improved version of its oversimplified linear and piece-wise linear counterparts. The authors in \cite{rev2,rev3} characterize the power conversion efficiency of the EH circuit as a second order polynomial and a rational function of the average input power, respectively. However, all these models fail to characterize the actual working principle of the harvesting circuit. On the other hand, the work in \cite{hparam} proposes a circuit-based realistic nonlinear EH model. This particular model not only relies on the EH circuit characteristics, but it also enables the design of waveforms that maximize the WPT efficiency.

The work in \cite{papr} demonstrates that in comparison to the conventional constant-envelope sinusoidal signals, certain waveforms with high peak-to-average-power-ratio (PAPR) provide higher harvested DC from the EH circuit. Based on this observation, some works, e.g. \cite{hparam,fhelps,npsk}, investigate the effect of transmitted waveforms and modulations on WPT and SWIPT. Due to the high PAPR of multisine waveforms, the work in \cite{hparam} proposes a multisine-based novel SWIPT architecture based on the superposition of multi-carrier unmodulated and modulated waveforms at the transmitter. The authors in \cite{fhelps}  develop a new signal design for WPT, which relies on multiple dumb antennas at the transmitter to induce fast fluctuations of the channel. The work in \cite{npsk} proposes an asymmetric modulation scheme specifically for SWIPT, that significantly enhances the rate-energy region as compared to its existing symmetric counterpart. Apart from the multisine waveforms, experimental studies demonstrate that due to their high PAPR, chaotic waveforms also are beneficial in terms of WPT efficiency \cite{chaosexp2}. The authors in \cite{wcl2} propose an analytical framework for continuous-time chaotic signals, which justifies the above observation.

Chaotic signals have been extensively used in the past decades for the purpose of wireless privacy and information security \cite{sec2}. However, in reality, digital operating platforms cannot support the continuous nature of the system variables, or signals and so, discretization of the continuous states or discrete approximations of the system have to be applied. Moreover, finding the solutions of differential functions costs computational capacity and hardware resources. In this context, as there is no heavy computational burden, the discrete-time chaotic system, namely the differential chaos shift keying (DCSK), is one of the most widely studied chaotic signal-based communication techniques \cite{bookcitingexample}. DCSK is a prominent benchmark in the class of noncoherent transmitted reference modulation techniques, which comprises of a reference and an identical/inverted replica of the reference depending on the data transmitted. The majority of the related works focus on the error performance of DCSK-based systems for various scenarios \cite{ch3,ch5,srdcsk}. The authors in \cite{ch3} propose an $M$-ary DCSK system, in which successive information bits are converted to a symbol and then transmitted by using the same modulation scheme. The work in \cite{ch5} investigates the performance of a cooperative diversity-aided DCSK-based system. Finally, to enhance the system data rate, the authors in \cite{srdcsk} propose a DCSK system with shorter symbol duration, namely, short reference DCSK (SR-DCSK).

In this direction, there are some existing works that exploit the benefits of both DCSK and WPT \cite{chaoswipt1,chaoswipt3,chaoswipt4}. The work in \cite{chaoswipt1} proposes an SR-DCSK based SWIPT architecture by using the time switching architecture to achieve higher data rate than conventional systems. The authors in \cite{chaoswipt3} investigate a chaotic carrier-index (CI) system for a basic SWIPT set-up to further reduce the energy consumption. Based on the transmission characteristics of index modulation, the proposed SWIPT scheme exploits the inactive carriers of CI-DCSK to wirelessly deliver energy by transmitting random noise-like signals. In \cite{chaoswipt4}, an adaptive link selection for buffer-aided relaying is investigated in a decode-and-forward relay-based DCSK-SWIPT architecture. Furthermore, by taking into account the decoding cost at the relays, two channel state information (CSI) unaware harvested energy-based link-selection schemes are proposed. Although these studies investigate DCSK-based SWIPT systems, they consider an impractical simplified linear model for harvesting. The works in \cite{jstsp} and \cite{dcskvtc} investigate DCSK-based waveform designs for WPT, where they consider the nonlinearities of the EH process to characterize the enhanced harvesting performance. However, the aspect of chaotic signal-based waveform designs for SWIPT and its analytical characterization have not been explored.

\begin{table*} [!t] 
\begin{center}
  \caption{Summary of notations.}
  \vspace{-2mm}
\resizebox{0.84\textwidth}{!}{%
{\renewcommand{\arraystretch}{2} 
  \begin{tabular}{|c|c||c|c|}
    \hline
    \bf{Notation} & \bf{Description} & \bf{Notation} & \bf{Description} \\
    \hline\hline
    $N$ & Number of receiver antennas &
    $M$ & Number of receiver antennas working in the IT mode\\
    \hline
    $r$ & Distance between transmitter and receiver & $K$ & Number of receiver antennas working in the EH mode\\
    \hline
    $a$ & Path-loss exponent &
    $L$ & Number of independent paths in the frequency-selective channel \\
    \hline
    $m$ & Channel fading parameter &
    $\alpha_{i,n}$ & $i$-th path, $n$-th antenna channel coefficient with $\mathbb{E}\{\alpha_{i,n}^2\}=\Omega_{i,n}$ \\
    \hline
    $\tau_i$ & $i$-th path delay & $x_{l,k}$ & $k$-th component, $l$-th transmission interval chaotic sequence
     \\
    \hline
    $\beta$ & Spreading factor & 
    $s_{l,k}$ & $k$-th component, $l$-th transmission interval transmitter output \\
    \hline
    $\phi$ & Reference length & $s_{l,k}^{R,n}$ & $n$-th antenna received signal for the transmit signal $s_{l,k}$ \\
    \hline
    $d_l$ & $l$-th information bit &
    $w_n$ & AWGN at $n$-th receiver antenna \\
    \hline
     $P_{\rm t}$ & Transmission power &
    $\mathbf{s}_l^{\rm PISO}$ & PISO shift register output for $l$-th information bit \\
    \hline
    $T_c$ & Chip duration & $s_l^{\rm AC}$ &  Analog correlator output for $l$-th information bit \\
    \hline
     $z_{\rm DC}$ & Harvested DC power & $k_2,k_4,R_{ant}$ & Harvesting circuit parameters \\
    \hline
    $\varepsilon_b$ & Transmitted bit energy &
     $\delta_{l,n}$ & Decision variable for $l$-th information bit at $n$-th antenna\\
    \hline
    $\gamma_0$ & SNR per bit at the receiver &
    $\delta_{l}$ & Decision variable for $l$-th information bit\\
    \hline    
  \end{tabular}
  }
   }
  \label{tab:notation}
\end{center}
\vspace{-2mm}
\end{table*}

Besides, note that the performance of conventional SWIPT systems is characterized by investigating the fundamental aspect of rate-energy trade-off \cite{swiptproc,swiptmimo,swipt2,hparam}. This is an effective way to quantify the performance of SWIPT systems and compare its various architectures. In this context, the Shannon capacity is used as the measure of data rate, which is valid if and only if the codeword is very long. Note that, the error probability or the bit error rate (BER) is not considered here because in the case of Shannon capacity, the corresponding codelength tends to infinity and thus, its associated BER asymptotically goes to zero. On the contrary, chaotic waveform-based codewords are finite length in nature and thus, investigating the conventional rate-energy trade-off is not enough for complete characterization of chaotic SWIPT systems. As a result, we investigate the BER-energy trade-off for such systems and also its impact on the transmit waveform design. Therefore, to summarize, the contributions of this paper are threefold.
\begin{itemize}
\item We introduce a novel DCSK-based single-input multiple-output (SIMO) SWIPT architecture by considering a generalized frequency selective Nakagami-$m$ fading scenario, where each receiver antenna switches between information transfer (IT) and EH modes. In this way, depending on the application's requirement, we accommodate the flexibility of prioritizing EH over IT or vice-versa or both.
\item The system BER is characterized as a function of the transmit waveform parameters, the number of antennas being utilized in the IT mode, and the channel parameters. Accordingly, we obtain the optimal reference length of the transmit waveform, which results in the minimum BER and we also derive the corresponding closed-form expression. Similarly, by taking into account the nonlinearities of the EH process, we characterize the harvested DC in the same way. The derived closed-form expressions, in both cases, are verified by extensive Monte Carlo simulations. They provide non-intuitive insights as to how the transmit waveform design and the system parameters affect the performance.
\item We investigate the BER-energy trade-off of the proposed chaotic architecture, where we introduce a novel achievable `success rate - harvested energy' region. Furthermore, we demonstrate that information and energy transfer are two conflicting tasks and hence, a particular waveform cannot be simultaneously optimal for both WPT and IT. As a result, we analyze two extreme scenarios, where all the receiver antennas are utilised for the IT or EH mode, respectively. Accordingly, we propose appropriate transmit waveform designs, which take into consideration the application specific acceptable BER or the harvested DC performance or both.
\end{itemize}
To the best of our knowledge, this is the first work that presents a complete analytical framework of chaotic waveform-based noncoherent signal design for SWIPT. It also takes into account the nonlinearities of the EH process and a realistic frequency selective wireless channel.

The rest of the paper is organized as follows. Section II introduces the proposed system architecture and our main assumptions. Section III presents the BER analysis, Section IV discusses the EH performance of the proposed framework, and Section V introduces the BER-energy trade-off. Finally, Section VI
presents the numerical results and Section VII concludes our work.

\textit{Notation}: $\mathbb{E}\{X\}$ and $\mathbb{P}\{X\}$ represent the expectation and the probability of $X$. $\Gamma(\cdot)$ is complete Gamma function and ${\rm erfc}(\cdot)$ denotes the complementary error function.

\section{System Model}
\begin{figure*}[!t]
 \centering\includegraphics[width=0.76\linewidth]{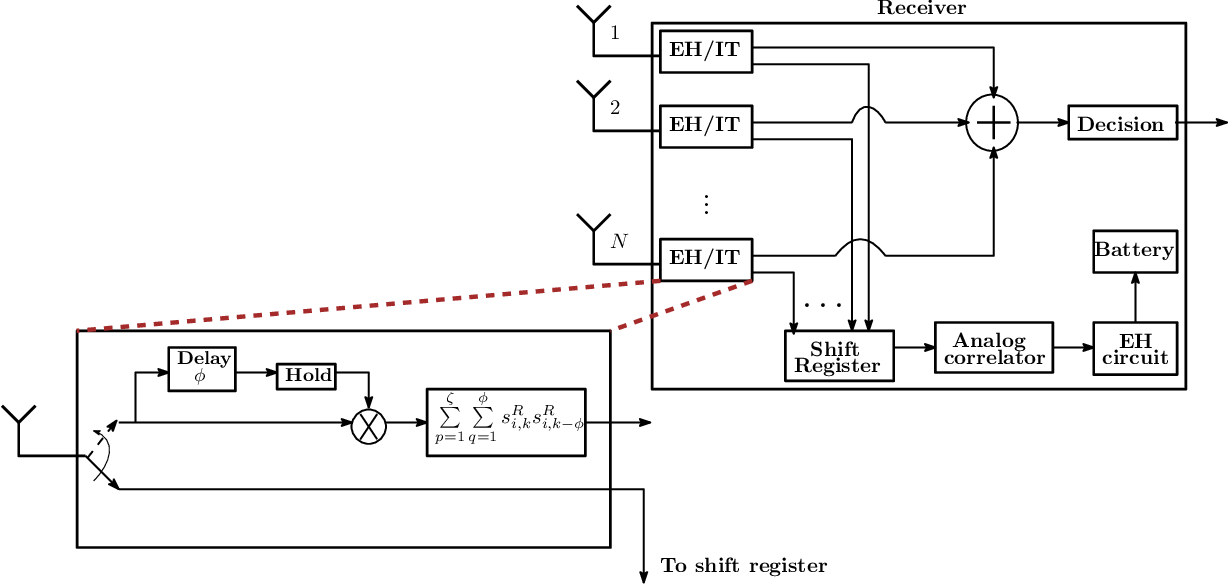}
\vspace{-2mm}
\caption{Chaotic signal-based proposed receiver architecture.}
\label{fig:model}
\vspace{-2mm}
\end{figure*}

In this section, we provide details of the considered system model; the main mathematical notations related to the system
model are summarized in Table \ref{tab:notation}. Specifically, as shown in Fig. \ref{fig:model}, we consider a SIMO SWIPT set-up, with a single antenna transmitter and an $N$ antenna receiver, where the transmitter employs a DCSK-based signal generator. At the receiver's side, we consider a multi-antenna architecture, where each antenna switches between IT or EH modes depending on the desired requirements (discussed in Section \ref{dscrptn}).

\subsection{Channel Model}  \label{chmodel}
We assume that the wireless link suffers from both large-scale path-loss effects and small-scale frequency-selective fading. Specifically, the received power is proportional to $r^{-a}$, where $r$ is the transmitter-receiver distance and $a>0$ denotes the path-loss exponent. It is worth noting that chaotic signals are generally wideband in nature \cite{2ray}. Hence, at an arbitrary receiver antenna, a frequency-selective channel is considered with $L$ independent paths following Nakagami-$m$ fading distribution \cite{papoulis}
\vspace{-1mm}
\begin{align}    \label{nakadef}
f_{\alpha_{i,n}}(\alpha)&=\frac{2m^m\alpha^{2m-1}e^{-\frac{m\alpha^2}{\Omega_{i,n}}}}{\Gamma(m)\Omega_{i,n}^m}, \nonumber \\
& \forall \:\:\alpha \geq 0, \:\: i=1,\dots,L \:\: \text{and} \:\: n=1,\dots,N,
\end{align}
where $\alpha_{i,n}$ is the channel coefficient corresponding to the $i$-th path at the $n$-th antenna with $\mathbb{E}\{\alpha_{i,n}^2\}=\Omega_{i,n}$ and $m \geq 1$ controls the severity of the amplitude fading. Furthermore, we assume that the sum of the corresponding power gains is one, i.e., $\sum\limits_{i=1}^L \Omega_{i,n}=1$ $\forall$ $n$ and consider identical channel statistics across all the $N$ antennas, i.e., $\Omega_{i,1}=\Omega_{i,2}=\dots=,\Omega_{i,N}=\Omega_i$ $\forall$ $i=1,\dots,L$.

\subsection{Chaotic signals}

Consider a DCSK signal, where a symbol is dependent on the previous one \cite{bookcitingexample} and different sets of chaotic sequences can be generated by using different initial conditions. Each transmitted bit is represented by two sets of chaotic signal samples, where the first set represents the reference, and the other conveys information. If the symbol $+1$ is to be transmitted, the data sample will be identical to the reference sample. Otherwise, an inverted version of the reference sample will be used as the data sample \cite{sec2}. During the $l$-th transmission interval, the transmitter output is
\vspace{-1mm}
\begin{align}  \label{sym}
s_{l,k}=\begin{cases} 
x_{l,k}, & k=2(l-1)\beta+1,\dots,(2l-1)\beta,\\
d_lx_{l,k-\beta}, & k=(2l-1)\beta+1,\dots,2l\beta,
\end{cases}&
\end{align}
where $d_l=\pm1$ is the information symbol, $x_{l,k}$ is the chaotic sequence used as the reference signal, and $x_{l,k-\beta}$ is its delayed version. If $\beta \in \mathbb{Z}^+$ is defined as the \textit{spreading factor},  $2\beta$ chaotic samples are used to spread each information bit \cite{sec2}. Furthermore, $x_{l,k}$ can be generated according to various existing chaotic maps. In this work, due to its good correlation properties, we consider the Chebyshev map $x_{k+1}=4x_k^3-3x_k$ for the generation of chaotic sequences \cite{bookcitingexample}.

\subsection{Proposed SR-DCSK-based Receiver System Design}  \label{srdcskdef}

\begin{figure*}[!t]
\centering\includegraphics[width=0.76\linewidth]{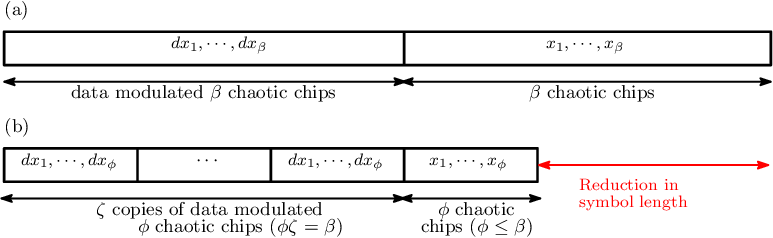}
\vspace{-2mm}
\caption{(a) DCSK frame, and (b) SR-DCSK frame.}
\vspace{-2mm}
\label{fig:sr}
\end{figure*}

Here, we extend the frame structure of \cite{srdcsk} in the context of a SIMO receiver architecture to integrate the SWIPT feature, where we select the EH/IT mode depending on the application's requirements. In SR-DCSK, an unmodulated chaotic component of length $\phi<\beta$ is considered, followed by $\zeta$ copies of its replica, multiplied with the information, such that $\beta=\zeta\phi$. Hence, the symbol duration is $\phi(1+\zeta)=\phi+\beta<2\beta$, i.e., the symbol is characterized by a sequence of $\phi+\beta$ samples of the chaotic basis signal, which is represented as
\vspace{-1mm}
\begin{align}  \label{symsr}
&s_{l,k} \nonumber \\
&=\!\begin{cases} 
x_{l,k}, & \! k=(l-1)(\beta+\phi)+1,\dots,(l-1)\beta+l\phi,\\
d_lx_{l,k-\phi}, & \!k=(l-1)\beta+l\phi+1,\dots,l(\beta+\phi),
\end{cases}&
\end{align}
\noindent An illustrative comparison of DCSK and SR-DCSK is demonstrated in Fig. \ref{fig:sr}; while Fig. \ref{fig:sr}(a) shows the conventional DCSK symbol structure, the short reference SR-DCSK is depicted in Fig. \ref{fig:sr}(b). It is important to note that the SR-DCSK is a technique proposed primarily for efficient IT with minimum BER. In contrast, we investigate the impact of having a shorter symbol duration on a multiple antenna receiver, where the antennas have the flexibility to switch between the IT and EH mode of operation. Hence, the received signal at the $n$-th antenna $(n=1,\dots,N)$ is
\vspace{-1mm}
\begin{equation}  \label{rcvd}
s^{R,n}_{l,k}=\sqrt{P_{\rm t}r^{-a}}\sum\limits_{i=1}^L \alpha_{i,n} s_{l,k-\tau_i}+w_n,
\end{equation}
where $P_{\rm t}$ is the transmission power, $\alpha_i$ and $s_{l,k-\tau_i}$ denote the channel coefficient and the delayed received signal corresponding to the $i$-th path, respectively, and $w_n$ is the additive white Gaussian noise (AWGN) at the $n$-th receiver antenna with zero mean and variance $\frac{N_0}{2}$ \cite{noise}. Moreover, practical scenarios suggest that the largest delay $\tau_L$ is significantly shorter compared to the reference length $\phi$ \cite{2ray}, i.e., $0 < \tau_L \ll \phi$, which results in
\vspace{-1mm}
\begin{equation}  \label{approxs}
s^{R,n}_{l,k} \approx \sqrt{P_{\rm t}r^{-a}}\sum\limits_{i=1}^L \alpha_{i,n} s_{l,k}+w_n.
\end{equation}

\subsection{Information and Power Transfer}  \label{dscrptn}

Suppose $M$ $(K)$ among the $N$ receiver antennas are selected for IT (EH) such that $M+K=N$. Therefore, during the period of the $l$-th transmitted information symbol, each of the EH/IT blocks associated with these $M$ antennas recover the chaotic component from the head of each SR-DCSK frame and perform $\zeta$ partial correlations over each block of $\phi$ samples. Finally, the obtained $\zeta$ values are summed up to be sent to the equal gain combiner (EGC) \cite{egc}, where the output from all the $M$ EH/IT blocks are combined and compared with a threshold to recover the actual transmitted data\footnote{An EGC is employed at the receiver to improve the error performance and is also motivated by its practical advantage of being easily implementable.}.

The $K$ antennas, which are selected for EH directly drive the received signal to a parallel-in serial-out (PISO) shift register \cite{piso}, whose output\footnote{The motivation of using the shift register is analogous to the working principle of a RF-combiner, where the received signal at all the $K$ antennas are combined together and then fed to the EH circuit \cite{comb}.} corresponding to the $l$-th transmitted information symbol can be expressed as
\vspace{-1mm}
\begin{align}  \label{srout}
\mathbf{s}_l^{\rm PISO}&=\left[ s^{R,1}_{l,1} s^{R,1}_{l,2} \cdots s^{R,1}_{l,\phi+\beta} \:\: s^{R,2}_{l,1} s^{R,2}_{l,2} \cdots s^{R,1}_{2,\phi+\beta} \:\: \cdots \right. \nonumber \\
& \qquad\qquad\qquad\qquad\cdots\left. \:\: s^{R,K}_{l,1} s^{R,K}_{l,2} \dots s^{R,K}_{l,\phi+\beta} \right].
\end{align}
In order to boost the EH performance, we consider an analog correlator that precedes the EH rectifier circuit \cite{jstsp}. An analog correlator essentially consists of a series of delay blocks, which result in signal integration over a specified period of time; an ideal $\psi$-bit analog correlator consists of $(\psi-1)$ number of delay blocks \cite{anaco2}. Therefore, as $K$ antennas are selected for EH with the transmit symbol duration being $\phi+\beta$, we consider $\psi=K(\phi+\beta)$. Accordingly, based on \eqref{srout}, the analog correlator output with respect to the $l$-th transmitted symbol is given by
\vspace{-1mm}
\begin{equation}  \label{corr}
s^{\rm AC}_l=\sum\limits_{n=1}^{K}\sum\limits_{k=1}^{\phi+\beta} s_{l,k}^{R,n}.
\end{equation}
Unlike an adder circuit \cite{ssmith}, the analog correlator generates an output, which is the sum of the delayed versions of the same signal \cite{anaco2}, i.e., the input signal is correlated with a delayed version of itself. The output of the analog correlator is fed to the EH circuit, which is basically a rectifier. Based on the nonlinearity of this circuit, the output DC current is expressed as \cite{hparam}
\vspace{-1mm}
\begin{equation} \label{brunoeh}
z_{\rm DC}=k_2R_{ant}\mathbb{E} \{ |s_l^{\rm AC}|^2 \}+k_4R_{ant}^2\mathbb{E} \{ |s_l^{\rm AC}|^4 \},
\end{equation}
where the parameters $k_2,k_4,$ and $R_{ant}$ are constants determined by the characteristics of the circuit. Note that the conventional linear model is a special case of this nonlinear model and can be obtained by considering only the first term in \eqref{brunoeh}. Hence, from \eqref{corr} and \eqref{brunoeh}, we obtain
\vspace{-1mm}
\begin{align} \label{zdef1}
z_{\rm DC}&=k_2R_{ant}\mathbb{E}\left\lbrace \left(  \sum\limits_{n=1}^{K}\sum\limits_{k=1}^{\phi+\beta} s_{l,k}^{R,n} \right) ^2 \right\rbrace \nonumber \\
& \qquad\qquad  + k_4R_{ant}^2\mathbb{E}\left\lbrace \left(  \sum\limits_{n=1}^{K}\sum\limits_{k=1}^{\phi+\beta} s_{l,k}^{R,n} \right) ^4 \right\rbrace,
\end{align}
where the expectation is taken over both the channel and the transmitted information symbol.

\section{Bit Error Rate Analysis}  \label{wit}

In this section we investigate the effect of SR-DCSK signals on the proposed receiver design, when $M$ $(\leq N)$ antennas are chosen for the IT mode. Based on the SR-DCSK frame design, if $T_c$ is the chip duration, the corresponding transmitted bit energy is given as
\vspace{-1mm}
\begin{equation}  \label{bnrgy}
\varepsilon_b=P_{\rm t} T_c \left( \phi+\beta \right) \mathbb{E}\{x_k^2\},
\end{equation}
where $x_k$ is the chaotic chip as defined in \eqref{symsr}. Accordingly, for the $l$-th transmitted symbol, the output from the EH/IT block associated with the $n$-th $(n=1,\dots,M)$ receiver antenna is given by
\begin{align}  \label{deltadef}
\delta_{l,n}&=T_c\sum\limits_{p=1}^{\zeta}\sum\limits_{q=0}^{\phi-1} \left( \sqrt{P_{\rm t}r^{-a}}\sum\limits_{i=1}^L \alpha_{i,n} x_{l,q}d_l+w_{p,q+\phi,n} \right) \nonumber \\
& \times\left( \sqrt{P_{\rm t}r^{-a}}\sum\limits_{i=1}^L \alpha_{i,n} x_{l,q}+w_{q,n} \right) \nonumber \\
&\overset{(a)}{=}\sum\limits_{p=1}^{\zeta}\sum\limits_{q=0}^{\phi-1} \left( P_{\rm t}r^{-a}\sum\limits_{i=1}^L \alpha_{i,n}^2 x_{l,q}^2d_l  \right. \nonumber \\
& \quad\left. + w_{q,n} \left( \sqrt{P_{\rm t}r^{-a}}\sum\limits_{i=1}^L \alpha_{i,n} x_{l,q}d_l \right) \right. \nonumber \\
& \quad\left. + w_{p,q+\phi,n} \left( \sqrt{P_{\rm t}r^{-a}}\sum\limits_{i=1}^L \alpha_{i,n} x_{l,q} \right) + w_{p,q+\phi,n}w_{q,n} \right) \nonumber \\
&=\zeta\phi P_{\rm t}r^{-a}\sum\limits_{i=1}^L \alpha_{i,n}^2 x_{l,q}^2d_l \nonumber \\
& + \zeta\sum\limits_{q=0}^{\phi-1}w_{q,n} \left( \sqrt{P_{\rm t}r^{-a}}\sum\limits_{i=1}^L \alpha_{i,n} x_{l,q}d_l \right) \nonumber \\
& + \sum\limits_{p=1}^{\zeta}\sum\limits_{q=0}^{\phi-1} \left( w_{p,q+\phi,n} \left( \sqrt{P_{\rm t}r^{-a}}\sum\limits_{i=1}^L \alpha_{i,n} x_{l,q} \right) \right. \nonumber \\
& + w_{p,q+\phi,n}w_{q,n} \Bigg),
\end{align}
where $(a)$ follows from considering $T_c=1$ for mathematical simplicity and by exploiting the low cross-correlation of two chaotic sequences, i.e.,
\vspace{-1mm}
\begin{equation}
\sum\limits_{q=0}^{\phi-1}x_{l,q_i}x_{l,q_j} \approx 0 \qquad \text{for} \quad i \neq j.
\end{equation}
Note that the noise component $w_{q,n}$ remains constant over all the $\zeta$ partial correlations. The evaluated $\delta_{l,n}$ from all the $M$ EH/IT blocks is sent to the EGC, in order to take an appropriate decision on the $l$-th transmitted data, i.e., the decision variable becomes $\delta_l=\sum\limits_{n=1}^{M}\delta_{l,n}$. For the sake of presentation, we define $\gamma_0=\frac{r^{-a}\varepsilon_b}{N_0}$. Now we analyse the system BER performance, as a function of $\delta_l$, which is given by the following theorem.
\begin{theorem}  \label{theoit1}
If $M$ $(\leq N)$ receiver antennas are utilized for the IT mode, the system's BER is
\vspace{-1mm}
\begin{align}  \label{berf}
&{\rm BER} \nonumber\\
&=\frac{1}{2}\int\limits_0^{\infty}{\rm erfc} \left(  \left[ \frac{M\left(\phi+\beta\right)^2}{2\beta\gamma_0^2\kappa^2}+\left(\frac{\zeta+1}{\zeta}\right)\left(\frac{\phi+\beta}{\phi\gamma_0\kappa}\right) \right]^{-\frac{1}{2}} \right)\nonumber\\
&\times f\left( \kappa \right){\rm d}\kappa,
\end{align}
where $\kappa=\sum\limits_{n=1}^{M}\sum\limits_{i=1}^L \alpha_{i,n}^2$, and $f\left( \kappa \right)$ is the probability distribution function of $\kappa$ given by
\vspace{-1mm}
\begin{equation}  \label{gpdf}
f(\kappa)=\frac{\left(mL \right)^{MmL}\kappa^{MmL-1}{\rm e}^{-\left(mL\kappa\right)}}{\Gamma \left( MmL \right) }, \qquad \kappa \geq 0.
\end{equation}
\end{theorem}
\begin{proof}
See Appendix \ref{app1}
\end{proof}

From the above theorem, we observe that the BER is a function of $M$, the SR-DCSK frame parameters, i.e., $\phi$ and $\beta$, as well as the channel parameters, i.e., $m$ and $\Omega_n$ $\forall$ $n=1,\dots,L$. Moreover, we also observe that an analytical closed-form expression of BER cannot be obtained in \eqref{berf} and numerical integration is the only way out. Hence, similar to the approach proposed in \cite{srdcsk}, we extend our analysis to the AWGN channel scenario, i.e., replace $\alpha_{1,z}=1$ and $\alpha_{i,n}=0$ $\forall$ $i \neq 1$, $n=1,\dots,M$ in $\kappa$ as defined in Theorem \ref{theoit1}. Accordingly, in the following theorem, we obtain the optimal reference length $\phi_{\rm opt}$ for a given set of $\beta,\gamma_0,$ and $M$, which minimizes the BER.
\begin{theorem}  \label{theoit2}
In the AWGN scenario, for given $\beta,\gamma_0,$ and $M,$ the system's BER is evaluated as ${\rm BER}=\frac{1}{2}{\rm erfc} \left(\sqrt{\frac{M}{\Lambda(\phi)}} \right)$, where $\Lambda(\phi)=\frac{\left(\phi+\beta\right)^2}{\beta\gamma_0}\left(\frac{1}{2\gamma_0}+\frac{1}{\phi}\right)$. Moreover, we obtain the minimum BER for the reference length $\phi_{\rm opt}=\frac{\gamma_0}{2} \left( \sqrt{1+\frac{4\beta}{\gamma_0}}-1 \right)$.
\end{theorem}
\begin{proof}
See Appendix \ref{app2}.
\end{proof}

\noindent Note that, as defined in Section \ref{srdcskdef}, we have $\beta=\zeta\phi$, where $\zeta$ is the number of replicas of the information modulated chaotic component of length $\phi$, i.e., $\frac{\beta}{\phi} \in \mathbb{Z}^+$. Hence, in cases where $\frac{\beta}{\phi_{\rm opt}}\notin  \mathbb{Z}^+$, we choose an appropriate $\phi$ closest to $\phi_{\rm opt}$ as the optimal $\phi$. It is interesting to observe from Theorem \ref{theoit2}, that unlike the minimum BER, the optimal reference length $\phi_{\rm opt}$ is independent of $M$. However, the monotonically decreasing nature of ${\rm erfc}(x)$ for $x \geq 0$ implies that the attainable minimum BER, for a given $\beta$ and $\gamma_0$, decreases with increasing $M$. Hence, to conclude, Theorem \ref{theoit2} demonstrates that, for a given $\beta,\gamma_0,$ and $M$, the BER is an unimodal function of $\phi \in [1,\beta]$. It  decreases monotonically when $\phi \in [1,\phi_{\rm opt})$, attains a minima at $\phi=\phi_{\rm opt}$, and starts to increase again when $\phi \in (\phi_{\rm opt},\beta]$. Hence, as our objective is to improve the system error performance, it is intuitive not to investigate the BER for range $\phi \in (\phi_{\rm opt},\beta]$, where there is no scope for improvement.

\section{Average Energy Harvesting}  \label{wpt}

In this section, we investigate the effect of SR-DCSK signals on the EH performance of the proposed receiver design, when $K$ $(\leq N)$ antennas are considered for EH. Specifically, we investigate the impact of the reference length $\phi$ on the harvested DC in terms of the spreading factor $\beta$ and the multipath fading wireless channel. By considering that the noise contribution to the harvested DC is negligible, we characterize the EH performance of the proposed receiver design by the following theorem.
\begin{theorem}  \label{theoeh1}
The harvested DC when $K(\leq N)$ antennas are being utilized for the EH mode is
\vspace{-1mm}
\begin{align}  \label{harvnp}
z_{\rm DC}&=\frac{\nu_1\Upsilon_1K^2\phi}{2}\left(  1+\zeta^2 \right) \nonumber \\
& + \frac{3\nu_2\Upsilon_2K^4}{8} \left( 1+6\zeta^2+\zeta^4 \right)\left( 2\phi^2-\phi \right),
\end{align}
\vspace{-2mm}
where $\nu_1 = r^{-a}k_2R_{ant}P_{\rm t}$, $\nu_2=r^{-2a}k_4R_{ant}^2P_{\rm t}^2$,
\begin{equation*}
\Upsilon_1=1+\frac{2}{m}\left(\frac{\Gamma(m+0.5)}{\Gamma(m)}\right)^2 \sum_{\mathclap{\substack{i_1,i_2=1\\i_1 \neq i_2}}}^L \sqrt{\Omega_{i_1}\Omega_{i_2}},
\end{equation*}
and
\vspace{-1mm}
\begin{equation*}
\Upsilon_2=\sum\limits_{i_1+i_2+\dots+i_L=4} \frac{4!}{i_1! \: i_2! \: \dots \: i_L!} \prod\limits_{j=1}^L  \frac{\Gamma(m+\frac{i_j}{2})}{\Gamma(m)}\!\!\left(\frac{\Omega_j}{m}\right)^{\frac{i_j}{2}}.
\end{equation*}
\end{theorem}
\begin{proof}
See Appendix \ref{app3}.
\end{proof}
Theorem \ref{theoeh1} provides a generalized closed-form expression for $z_{\rm DC}$ in terms of the frequency selective channel parameters, i.e.,  $m$ and $\Omega_i$ $\forall$ $i=1,\dots,L$, and also the SR-DCSK related parameters, namely, the reference length $\phi$ and $\beta$ $(\beta=\zeta\phi)$. In this case, $z_{\rm DC}$ corresponding to a flat fading scenario and no-fading environment can also be obtained as special cases. Towards this direction, we state the following proposition.
\begin{prop}  \label{theoeh2}
For a flat fading channel, the harvested DC is 
\vspace{-1mm}
\begin{align}  \label{flat}
z_{\rm DC,FF}&=\frac{\nu_1K^2\phi}{2}\left(  1+\zeta^2 \right) \nonumber\\
& \!\!\!\!\!\!+ \frac{3\nu_2K^4}{8} \left( \frac{1+m}{m} \right) \left( 1+6\zeta^2+\zeta^4 \right)\left( 2\phi^2-\phi \right)
\end{align}
while for a no-fading scenario, we obtain
\vspace{-1mm}
\begin{align}  \label{nof}
z_{\rm DC,NF}&=\frac{\nu_1K^2\phi}{2}\left(  1+\zeta^2 \right) \nonumber\\
& + \frac{3\nu_2K^4}{8} \left( 1+6\zeta^2+\zeta^4 \right)\left( 2\phi^2-\phi \right).
\end{align}
\end{prop}
\begin{proof}
See Appendix \ref{app5}.
\end{proof}
From Theorem \ref{theoeh1} and Proposition \ref{theoeh2}, we observe the impact of the number of receiver antennas being utilized in  the EH mode on the harvested DC, for a given set of system parameters. Note that $z_{\rm DC}$ increases with $K$ and $\zeta$ in the order of $K^4\zeta^4 \equiv \left(\frac{K}{\phi}\right)^4$, for a given $\beta$. This observation corroborates the claims made in \cite{jstsp} regarding the beneficial role of DCSK-based signal design in the context of WPT. Moreover, it can also be seen from \eqref{flat} that $z_{\rm DC, FF}$ decreases with increasing $m$, i.e., fading enhances EH, which further endorses the investigation conducted in \cite{jstsp}. For a given fading scenario, if the system parameters, i.e., $\nu_1,\nu_2,$ and $K$, are assumed to be constant, the harvested DC from Theorem \ref{theoeh1} can be expressed in terms of $\phi$ and $\zeta$ as
\vspace{-1mm}
\begin{equation}
z_{\rm DC}=\varphi_1\phi\left(  1+\zeta^2 \right) + \varphi_2 \left( 1+6\zeta^2+\zeta^4 \right)\left( 2\phi^2-\phi \right),
\end{equation}
where we define the constants $\varphi_1=\frac{\nu_1\Upsilon_1K^2}{2}$ and $\varphi_2=\frac{3\nu_2\Upsilon_2K^4}{8}$. By using $\beta=\phi\zeta$, where $\beta$ is fixed, $z_{\rm DC}$ can be further rewritten  in terms of $\phi$ as
\vspace{-1mm}
\begin{equation}
z_{\rm DC}(\phi)=\frac{\varphi_1}{\phi} \left( \beta^2+\phi^2 \right) + \frac{\varphi_2}{\phi^3}\left( \phi^4+6\beta^2\phi^2+\beta^4 \right)\left( 2\phi-1 \right).
\end{equation}
By definition, we have $\phi \geq 1$. Hence, the maximum $z_{\rm DC}$ is obtained for $\phi=1$, i.e.,
\vspace{-1mm}
\begin{equation}  \label{zdcmax}
z_{\rm DC}^{\max}=\varphi_1 \left( 1+\beta^2 \right) + \varphi_2\left( 1+6\beta^2+\beta^4 \right),
\end{equation}
for a symbol length  of $\beta+1$. Therefore, the WPT optimal DCSK-based waveform is as follows.
\vspace{-1mm}
\begin{align}  \label{optsym}
&s_{l,k} \nonumber \\
&\!\!=\!\!\begin{cases} 
\!\!x_{l,k}, & \!\!\! k=(l-1)(\beta+1)+1,\\
\!\!d_lx_{l,(l-1)(\beta+1)+1}, & \!\!\! k=(l-1)(\beta+1)+2,\dots,l(\beta+1).
\end{cases}&
\end{align}
Accordingly, the optimal $z_{\rm DC}$ is obtained by replacing $\phi=1$ and $\zeta=\beta$ in \eqref{harvnp}, i.e.,
\vspace{-1mm}
\begin{equation}  \label{zdcopt}
z_{\rm DC}^{\rm opt}=\frac{\nu_1\Upsilon_1K^2}{2}\left(  \beta^2+1 \right)  + \frac{3\nu_2\Upsilon_2K^4}{8} \left( \beta^4+6\beta^2+1 \right).
\end{equation}
Note that, the work in \cite{jstsp} and \cite{dcskvtc} dealt with DCSK-based waveforms for WPT in the context of a single antenna receiver. Hence, for the sake of completeness, we also provide $z_{\rm DC}$ corresponding to identical scenarios in the following corollary.

\begin{corr}
With a single antenna at the receiver being utilized in the EH mode, we obtain
\vspace{-1mm}
\begin{equation}
z_{\rm DC,FF}=\frac{\nu_1}{2}\left(1+\beta^2 \right)+\frac{3\nu_2(1+m)}{8m}\left( 1+6\beta^2+\beta^4 \right),
\end{equation}
for a flat fading scenario and
\vspace{-1mm}
\begin{align}
z_{\rm DC,FS}&=\frac{\nu_1}{2}\left(1+\beta^2 \right)\!\left( 1+\frac{2}{m}\left(\frac{\Gamma(m+0.5)}{\Gamma(m)}\right)^2 \sqrt{\Omega_1\Omega_2} \right) \nonumber \\
&\!\!\!\!\!\!\!\!\!\!\!\!\!\!\!\!+\frac{3\nu_2}{8} \left( 1+6\beta^2+\beta^4 \right)\left( \left( \Omega_1^2+\Omega_2^2 \right) \left( \frac{m+1}{m} \right)+6\Omega_1\Omega_2 \right.\nonumber\\
&\left.\!\!\!\!\!\!\!\!\!\!\!\!\!\!\!\!+4\frac{\Gamma
(1.5+m) \Gamma (0.5+m)}{m^2 \Gamma(m)^2} \left( \Omega_1^{0.5}\Omega_2^{1.5}+\Omega_1^{1.5}\Omega_2^{0.5} \right)\right),
\end{align}
for a two-ray frequency selective environment.
\end{corr}
\noindent The above corollary follows directly from Theorem \ref{theoeh1}, where $z_{\rm DC,FF}$ is obtained by replacing $K=1,\phi=1,\Omega_1=1,$ and $\Omega_i=0$ $\forall$ $i>1$ in \eqref{harvnp}. Furthermore, $z_{\rm DC,FS}$ refers to the special case corresponding to \eqref{harvnp} with $K=1,\phi=1,$ and $\Omega_i=0$ $\forall$ $i>2$, respectively. This implies that the receiver design proposed in this work is much more generalized as compared to the former two.

\section{Waveform Design and BER-Energy Trade-off Characterization}  \label{region}
We discuss the joint BER performance and harvested energy of the considered chaotic multi-antenna receiver. We characterize the receiver performance in terms of the achievable success rate - harvested energy $({\rm SR}-z_{\rm DC})$ region, where we define `success rate' as ${\rm SR}=1-{\rm BER}$. In the previous section, it is observed that for a given spreading factor $\beta$, the maximum harvested DC is obtained at $\phi=1$ and accordingly, a WPT-optimal waveform design is proposed in \eqref{optsym}. On the other hand, Section \ref{wit} demonstrates that the optimal data transmission strategies for having maximum ${\rm SR}$ leads to different values of $\phi$; specifically, ${\rm SR}$ increases monotonically when $\phi \in [1,\phi_{\rm opt}]$ (discussed in Section \ref{wit}). The above observation is justified as follows.

Information and energy contents are two contrasting criteria as far as the design of a waveform is concerned. While randomness by means of correlation in the transmitted signal is beneficial for obtaining higher $z_{\rm DC}$, it comes at the cost of degraded ${\rm SR}$, and vice-versa. In this context, for a given $\beta$, increasing $\phi$ results in diminishing correlation within a SR-DCSK frame, which translates to a deteriorating $z_{\rm DC}$ and an enhanced ${\rm SR}$ performance, respectively. Note that, this observation is in line with the claims made in \cite{corrl}, where the contrasting roles of correlation in both information and energy transfers are investigated. Thus, we arrive at the following question: what is the optimal SR-DCSK frame structure for simultaneous noncoherent information and energy transfer? Moreover, for a fixed transmission power, we also decide on the number of antennas at the receiver, which are used for information and energy extraction from the received signal, respectively. In this context, we propose to define the ${\rm SR}-z_{\rm DC}$ region as follows. Without any loss of generality, by assuming that the transmitter sends SR-DCSK frames of length $\beta+\phi$, the ${\rm SR}-z_{\rm DC}$ region is defined as
\begin{align}  \label{regdef}
&\mathcal{C}_{{\rm SR}-z_{\rm DC}}\left( \phi,M,K \right) \nonumber\\
& \!\!= \left\lbrace \!\! \left( {\rm SR}^{\rm M}, z_{\rm DC}^{\rm M} \right)\!:\!\frac{1}{2}{\rm erfc} \left(\!\! \sqrt{\frac{M}{\Lambda(\phi)}} \right)\!\! \leq 1-{\rm SR}^{\rm M},  z_{\rm DC} \geq z_{\rm DC}^{\rm M}\!\! \right\rbrace, 
\end{align}
where $z_{\rm DC}$ is obtained from Theorem \ref{theoeh1} based on the channel conditions and for a suitable set of $\phi$ and $K$, such that $K+M=N$. We also observe that $\phi$ is a crucial factor in this ${\rm SR}-z_{\rm DC}$ region characterization, i.e., the chaotic repetition mechanism forms an integral part of the proposed system design. This is because, we have $\beta=\zeta\phi$, where $\zeta$ is the number of replicas of the information modulated chaotic component of length $\phi$, i.e., $\frac{\beta}{\phi} \in \mathbb{Z}^+$. Furthermore, ${\rm SR}^{\rm M}$ and $z_{\rm DC}^{\rm M}$ denotes the minimum acceptable success rate and harvested DC, respectively.

Note that the parameter $\beta$ is fixed for the ${\rm SR}-z_{\rm DC}$ region defined in \eqref{regdef} and that it has three control parameters, namely, $\phi,M,$ and $K$, respectively. To obtain more analytical insights, we consider the following extreme scenarios.
\begin{enumerate}
\item Achievable ${\rm SR}$ when $M=N$, and
\item Harvested DC corresponding to $K=N$.
\end{enumerate}

\subsection{Achievable ${\rm SR}$ when $M=N$.}
Based on Theorem \ref{theoit2}, for a given $\beta,\gamma_0,$ and $N$, the maximum achievable ${\rm SR}$ is obtained as
\vspace{-1mm}
\begin{equation}
{\rm SR}=1-\frac{1}{2}{\rm erfc} \left(\sqrt{\frac{N}{\Lambda(\phi_{\rm opt})}} \right),
\end{equation}
where we have $\Lambda(\phi_{\rm opt})=\frac{\left(\phi_{\rm opt}+\beta\right)^2}{\beta\gamma_0}\left(\frac{1}{2\gamma_0}+\frac{1}{\phi_{\rm opt}}\right)$ and the corresponding optimal reference length is $\phi_{\rm opt}=\frac{\gamma_0}{2} \left( \sqrt{1+\frac{4\beta}{\gamma_0}}-1 \right)$. Moreover, we observe that $\phi_{\rm opt}$ is a function of $\beta$ and $\gamma_0$, whereas ${\rm SR}$ is a function of $\phi_{\rm opt}$ and $N$. Hence, for a specific $\beta$ and $\gamma_0$, if we are interested in designing a SR-DCSK based $N$ antenna receiver, trivial algebraic manipulations of the analytical expression of ${\rm SR}$ as obtained above leads to
\vspace{-1mm}
\begin{equation}
N=\left( {\rm erfc}^{-1} \left(  2 \left(1-{\rm SR} \right)\right) \right)^2 \Lambda(\phi_{\rm opt}).
\end{equation}

\subsection{Harvested DC corresponding to $K=N$.}
This scenario implies that all the $N$ antennas are being solely used for power and not IT. As a result, in this case, transmission of SR-DCSK waveforms, as defined in \eqref{symsr}, implies unnecessary wastage of resources, i.e., data. Thus, here we consider the case of using an unmodulated, deterministic chaotic waveform for WPT. Specifically, the transmitted symbol is characterized by a sequence of $\beta+1$ chaotic samples, which is represented as
\vspace{-1mm}
\begin{equation}  \label{onlychaos}
s_{l,k}=x_{l,k}, \qquad k=(l-1)(\beta+1)+1,\dots,l(\beta+1).
\end{equation}
It is worth noting that, as defined in \eqref{optsym}, for a given $\beta$, it is essential to have a symbol of length $\beta+1$ for WPT-optimal SR-DCSK but no such constraint is needed here. However, we consider a symbol length $\beta+1$ for a fair comparison of the EH performance. Now, we state the following theorem, where we characterize the harvested DC for unmodulated chaotic waveforms.
\begin{theorem}  \label{onlychaosth}
The harvested DC, in the case of unmodulated waveform transmission, is given by
\vspace{-1mm}
\begin{equation}  \label{zdcochaos}
z_{\rm DC}^{\rm UM}=\frac{\nu_1\Upsilon_1N^2}{2}\left(  \beta+1 \right) + \frac{3\nu_2\Upsilon_2N^4}{8} \left( 2\beta^2+3\beta+1 \right),
\end{equation}
where $\nu_1,\nu_2,\Upsilon_1,$ and $\Upsilon_2$ are as defined in Theorem \ref{theoeh1}.
\end{theorem}
\begin{proof}
The proof is similar to the one of Theorem \ref{theoeh1}. The only point that we need to consider here is that based on \eqref{onlychaos}, in this case, we have $\sum\limits_{k=1}^{\beta+1} s_{l,k}=\sum\limits_{k=1}^{\beta+1}x_{l,k}$.
\end{proof}

\noindent We observe from \eqref{zdcopt} and \eqref{zdcochaos} that, for identical set of system and channel parameters, we have $z_{\rm DC}^{\rm UM}<z_{\rm DC}^{\rm opt}$ when $K=N$. The reason behind this observation is attributed to the importance of correlation in signal design for enhanced EH performance (discussed in Section \ref{region}). Specifically, in \eqref{optsym}, only the first symbol is chaotic, followed by $\beta$ copies of their replica or inverted replica, i.e., there is a high degree of `similarity' or correlation. On the contrary, no such correlation exists among the $\beta+1$ chaotic components that constitutes the transmitted symbol in \eqref{onlychaos}. As a result, when we are solely interested in EH, without any wastage of resources, we can exploit the maximum correlation possible by consecutively placing identical chaotic elements in the transmitted chaotic frame. Accordingly, during the $l$-th transmission interval, the output of the transmitter is
\vspace{-1mm}
\begin{align}  \label{symwpt}
&s_{l,k} \nonumber \\
&=\!\!\begin{cases} 
x_{l,k}, & \!\!k=(l-1)(\beta+1)+1,\\
x_{l,(l-1)(\beta+1)+1}, & \!\!k=(l-1)(\beta+1)+2,\dots,l(\beta+1).
\end{cases}&
\end{align}
Next, we characterize the EH performance of \eqref{symwpt} in the following theorem.

\begin{theorem}  \label{theoeh3}
When all the receiver antennas are being utilized in the EH mode, the frame structure defined in \eqref{symwpt} results in the harvested DC given by
\vspace{-1mm}
\begin{equation}  \label{harvwpt}
z_{\rm DC}^{\rm PT}=\frac{\nu_1\Upsilon_1N^2}{2}\left(  \beta+1 \right)^2 + \frac{3\nu_2\Upsilon_2N^4}{8} \left( \beta+1 \right)^4,
\end{equation}
where $\nu_1,\nu_2,\Upsilon_1,$ and $\Upsilon_2$ are as defined in \eqref{harvnp}, respectively.
\end{theorem}
\begin{proof}
The proof follows similarly as the one of Theorem \ref{theoeh1}. The main point that we need to consider here is that based on \eqref{symwpt}, in this case, we have $\sum\limits_{k=1}^{\beta+1} s_{l,k}=(\beta+1)x_{l,k}$.
\end{proof}
\noindent To obtain analytical insights into the harvested DC when $K=N$, we consider a flat fading channel assumption, as given in the following corollary.
\begin{corr}
By considering the limiting case of a flat fading scenario, the harvested DC is
\vspace{-1mm}
\begin{equation}
z_{\rm DC}^{\rm PT}=\frac{\nu_1N^2}{2}\left(  \beta+1 \right)^2 + \frac{3\nu_2N^4}{8} \left( \frac{1+m}{m} \right) \left( \beta+1 \right)^4.
\end{equation}
\end{corr}
\noindent The above corollary follows directly by replacing $\Omega_1=1$ and $\Omega_i=0$ $\forall$ $i>1$. Moreover, here also, we also observe the beneficial role of fading in the context of WPT, as discussed earlier in Section \ref{wpt}.

\begin{table*} [!t] 
\begin{center}
  \caption{Summary of results when $K=N$.}
\vspace{-2mm}
\resizebox{0.96\textwidth}{!}{%
{\renewcommand{\arraystretch}{2} 
  \begin{NiceTabular}{|c||c|c|}[hvlines,cell-space-limits=3pt] 
    \hline
    \bf{Strategy} & \bf{Transmit Signal Design} & \bf{Harvested DC}\\
    \hline\hline
    $z_{\rm DC}^{\rm UM}$   & $s_{l,k}=x_{l,k}, \qquad k=(l-1)(\beta+1)+1,\dots,l(\beta+1).$ & $\frac{\nu_1\Upsilon_1N^2}{2}\left(  \beta+1 \right) + \frac{3\nu_2\Upsilon_2N^4}{8} \left( 2\beta^2+3\beta+1 \right)$\\
    \hline
    $z_{\rm DC}^{\rm opt}$  & $\begin{aligned}[t]
s_{l,k}=\begin{cases} 
x_{l,k}, & k=(l-1)(\beta+1)+1,\\
d_lx_{l,(l-1)(\beta+1)+1}, & k=(l-1)(\beta+1)+2,\dots,l(\beta+1).
\end{cases}&
\end{aligned}$ & $\frac{\nu_1\Upsilon_1N^2}{2}\left(  \beta^2+1 \right)  + \frac{3\nu_2\Upsilon_2N^4}{8} \left( \beta^4+6\beta^2+1 \right)$ \\
    \hline
    $z_{\rm DC}^{\rm PT}$  &$\begin{aligned}[t]
s_{l,k}=\begin{cases} 
x_{l,k}, & k=(l-1)(\beta+1)+1,\\
x_{l,(l-1)(\beta+1)+1}, & k=(l-1)(\beta+1)+2,\dots,l(\beta+1).
\end{cases}&
\end{aligned}$ & $\frac{\nu_1\Upsilon_1N^2}{2}\left(  \beta+1 \right)^2 + \frac{3\nu_2\Upsilon_2N^4}{8} \left( \beta+1 \right)^4$\\
\hline
  \end{NiceTabular}
  }
  }
  \label{tab:summ}
\end{center}
\vspace{-2mm}
\end{table*}

For the $K=N$ scenario, Table \ref{tab:summ} summarizes the various transmit signal designs (in the increasing order of correlation) and their corresponding $z_{\rm DC}$. As discussed in Section \ref{region}, here we observe that increasing correlation significantly enhances the EH performance of a waveform, i.e., $z_{\rm DC}^{\rm UM}<z_{\rm DC}^{\rm opt}<z_{\rm DC}^{\rm PT}$, both in terms of correlation and harvested DC. Moreover, if we have a closer look at the analytical results, an interesting insight can be obtained. In particular, we characterize the performance gap $\xi_1=z_{\rm DC}^{\rm PT}-z_{\rm DC}^{\rm UM}$ and $\xi_2=z_{\rm DC}^{\rm PT}-z_{\rm DC}^{\rm opt}$, where we have
\vspace{-1mm}
\begin{align}  \label{pgap}
\xi_1&=\!\frac{\nu_1\Upsilon_1N^2}{2}\!\left(  \beta^2+\beta \right) + \frac{3\nu_2\Upsilon_2N^4}{8} \!\left( \beta^4+4\beta^3+4\beta^2+\beta \right), \nonumber \\
\xi_2&=\!\nu_1\Upsilon_1N^2\beta + \frac{3\nu_2\Upsilon_2N^4}{2} \left( \beta^3+\beta \right).
\end{align}
As we know that $\beta \in \mathbb{Z}^+$, by comparing $\xi_1$ and $\xi_2$, we have $\xi_1>\xi_2$, with all other system and channel parameters remaining constant. This observation is explained as follows. Based on the properties of chaotic waveforms, there is no correlation in the transmit signal design corresponding to $z_{\rm DC}^{\rm UM}$. On the other hand, $z_{\rm DC}^{\rm opt}$ does exhibit some degree of correlation with limited randomness due to having $d_l=\pm 1$. Finally, the signal design of $z_{\rm DC}^{\rm PT}$ is absolutely correlated with $\beta+1$ copies identical chaotic components being consecutively placed to form a transmit symbol. Note that, if we remove the randomness from $z_{\rm DC}^{\rm opt}$, i.e., if $d_l$ is not `random' any more, $z_{\rm DC}^{\rm opt}$ gets transformed to $z_{\rm DC}^{\rm PT}$.

Here we analytically characterize the performance gain of the proposed waveform over and above the conventional DCSK for SWIPT. Specifically, let $\left\lbrace {\rm SR_P},z_{\rm DC,P} \right\rbrace $ and $\left\lbrace {\rm SR_D},z_{\rm DC,D} \right\rbrace $ denote the $\left\lbrace {\rm success \:\: rate}, {\rm harvested \:\: DC} \right\rbrace $ pair of the proposed and conventional DCSK, respectively. From Theorem \ref{theoit2} , for given $\beta,\gamma_0,$ and $M,$  we obtain
\vspace{-1mm}
\begin{equation}
{\rm SR_P}=1-\frac{1}{2}{\rm erfc} \left(\sqrt{\dfrac{M}{\Lambda(\phi)}} \right),
\end{equation}
where $\Lambda(\phi)=\frac{\left(\phi+\beta\right)^2}{\beta\gamma_0}\left(\frac{1}{2\gamma_0}+\frac{1}{\phi}\right)$ and the corresponding maximum SR is attained for the reference length $\phi_{\rm opt}=\frac{\gamma_0}{2} \left( \sqrt{1+\frac{4\beta}{\gamma_0}}-1 \right)$. On the other hand, for the conventional DCSK we have $\phi=\beta$, i.e., ${\rm SR_D}=1-\dfrac{1}{2}{\rm erfc} \left(\sqrt{\dfrac{M}{\Lambda(\beta)}} \right)$. Followed by some algebraic manipulations and the fact that ${\rm erfc}(x)$ is monotonically decreasing for $x \geq 0$, we obtain
\vspace{-1mm}
\begin{equation}  \label{src}
{\rm SR_D}<{\rm SR_P}.
\end{equation}
Furthermore, based on $\phi$ and Theorem \ref{theoeh1}, we obtain the harvested DC as
\vspace{-1mm}
\begin{align}
z_{\rm DC,P}&=\frac{\nu_1\Upsilon_1K^2\phi}{2}\left(  1+\zeta^2 \right) \nonumber \\
& + \frac{3\nu_2\Upsilon_2K^4}{8} \left( 1+6\zeta^2+\zeta^4 \right)\left( 2\phi^2-\phi \right),
\end{align}
where $\nu_1,\nu_2,\Upsilon_1,\Upsilon_2$ are system parameters and $K$ is the number of receiver antennas being utilized in the EH mode. As discussed previously, we have $\phi=\beta$ for conventional DCSK, i.e., $\zeta=1$, which results in the corresponding
\vspace{-1mm}
\begin{equation}
z_{\rm DC,D}=\nu_1\Upsilon_1K^2\beta + 3\nu_2\Upsilon_2K^4 \left( 2\beta^2-\beta \right).
\end{equation}
\noindent By comparing the second and fourth order terms of the EH process
separately, we obtain
\vspace{-1mm}
\begin{equation}  \label{zdcc}
z_{\rm DC,D}<z_{\rm DC,P}.
\end{equation}
By combining \eqref{src} and \eqref{zdcc}, we observe that the SR-$z_{\rm DC}$ region corresponding to the proposed waveform shrinks, when the conventional DCSK is used as the existing benchmark.

\section{Numerical Results}
We provide numerical results to demonstrate the performance of the proposed multi-antenna SWIPT receiver design and validate our theoretical analysis. Without any loss of generality, we consider a transmission power of $P_t=30$ dBm, a Tx-Rx distance $r=20$ m, and a pathloss exponent $a=4$ \cite{ploss}. The parameters considered for the EH model are taken as $k_2=0.0034,k_4=0.3829,$ and $R_{ant}=50$ $\Omega$ \cite{hparam}. Unless otherwise stated, we consider a frequency selective scenario with two independent paths for the purpose of presenting our results. Finally, note that, the considered set of parameter values is used for the sake of presentation. A different set of values will affect the performance but will lead to similar observations.

\begin{figure}[!t]\centering\includegraphics[width=\linewidth]{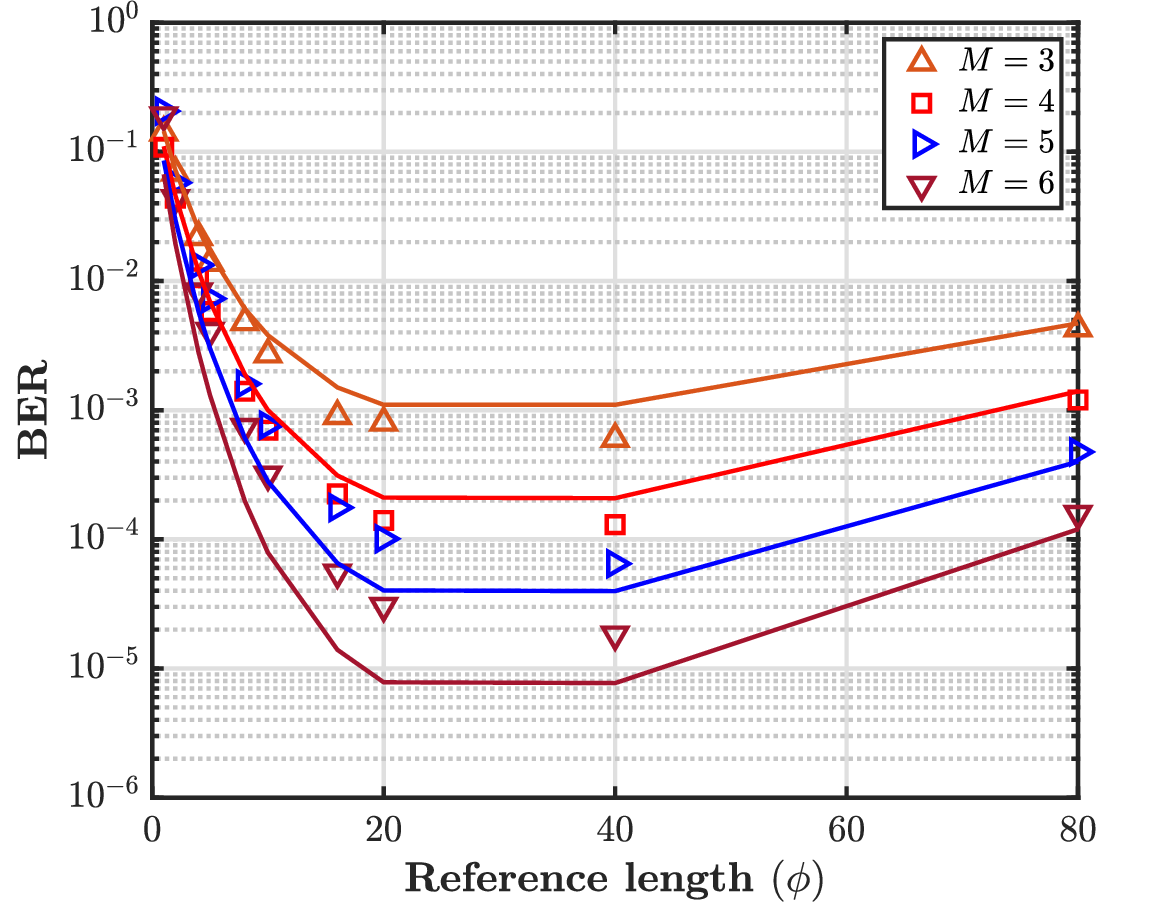}
\vspace{-4mm}
\caption{Impact of $M$ on BER; $\gamma_0=12$ dB and $\beta=80$; lines and markers correspond to analytical and simulation results, respectively.}
\vspace{-2mm}
\label{fig:optphi}
\end{figure}

Fig. \ref{fig:optphi} demonstrates the impact of $M$ on the BER performance of the proposed receiver design with $\beta=80$ in an AWGN scenario, as a function of the reference length $\phi$. The figure illustrates the BER performance with respect to the reference length $\phi$, for $M=3,4,5$ and $6$, respectively. Note that although the optimal reference length $\phi_{\rm opt}$, for a given $\gamma_0$ and $\beta$, is same $\forall$ $M$, ${\rm BER}_{\rm opt}$ decreases rapidly with increasing $M$ (Theorem \ref{theoit2}). Moreover, when $M$ is fixed, BER initially decreases with increasing $\phi$, attains minimum value at $\phi=\phi_{\rm opt}$ and starts increasing again. Furthermore, we observe that there is gap in between the simulation (markers) and the theoretical (lines) results for lower values of $\phi$, which gradually decreases as $\phi$ increases. The reason behind this observation is attributed to the Gaussian approximation in the theoretical derivation of BER.

\begin{figure}[!t]\centering\includegraphics[width=\linewidth]{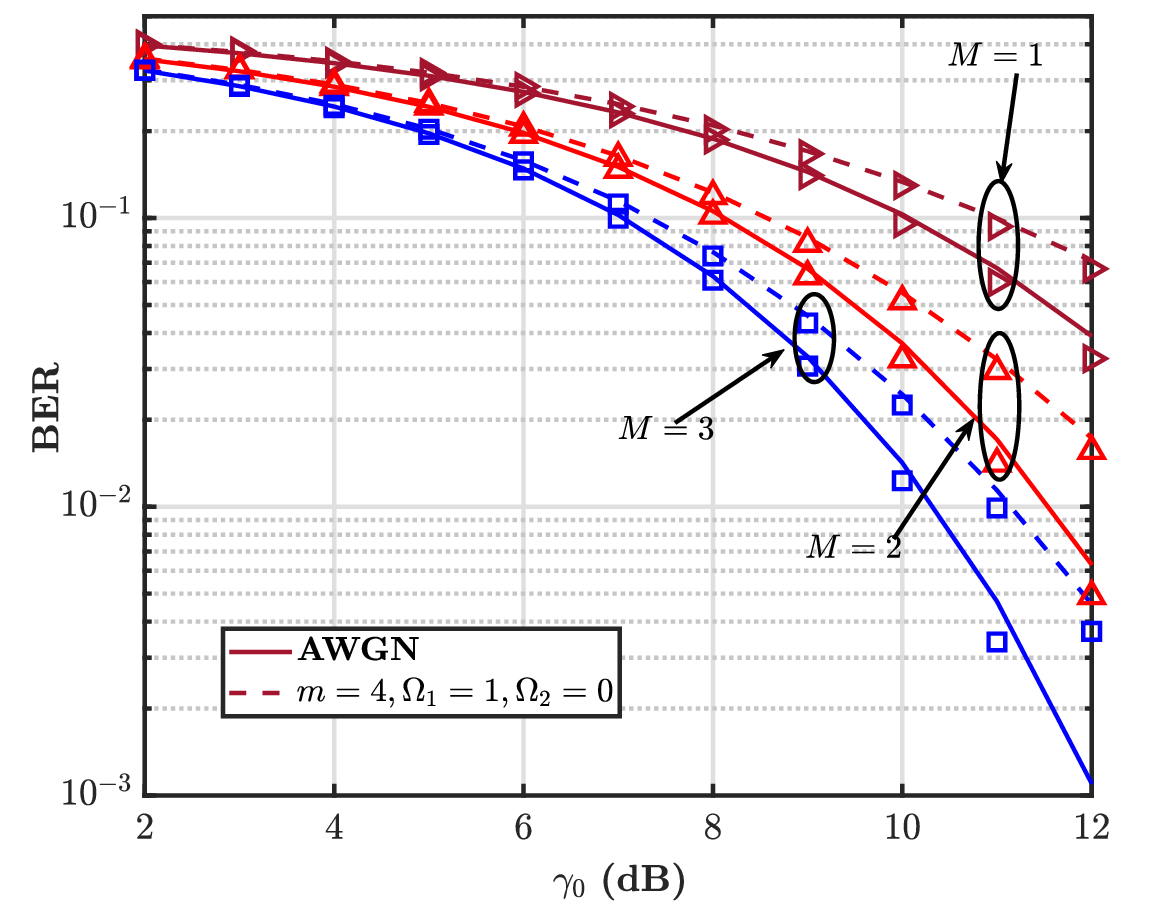}
\vspace{-4mm}
\caption{Impact of $\gamma_0$ on BER; $\beta=80$ and $\phi=20$; lines and markers correspond to analytical and simulation results, respectively.}
\vspace{-2mm}
\label{fig:bergamma}
\end{figure} 

In Fig. \ref{fig:bergamma}, we show the achieved BER performance against $\gamma_0$ for different fading scenarios with $M=1,2,$ and $3$, respectively. Clearly, an increase in $M$ improves the BER since the receive diversity increases with $M$. It is obvious from the figure, that irrespective of the channel conditions,  significant improvement in BER can be obtained, at moderate to high $\gamma_0$, with increasing $M$. Specifically, at $\gamma_0=12$ dB in an AWGN scenario, the proposed receiver design results in approximately $12$ dB performance improvement with $M=3$, when compared against $M=1$. On the other hand, we observe that fading degrades the BER performance for identical set of system parameters. However, this is expected since fading implies that random fluctuations are getting introduced to the transmitted waveform on its way to the receiver. On the contrary, it is important to note that fading is beneficial for WPT (Section \ref{wpt}).

\begin{figure}[!t]\centering\includegraphics[width=\linewidth]{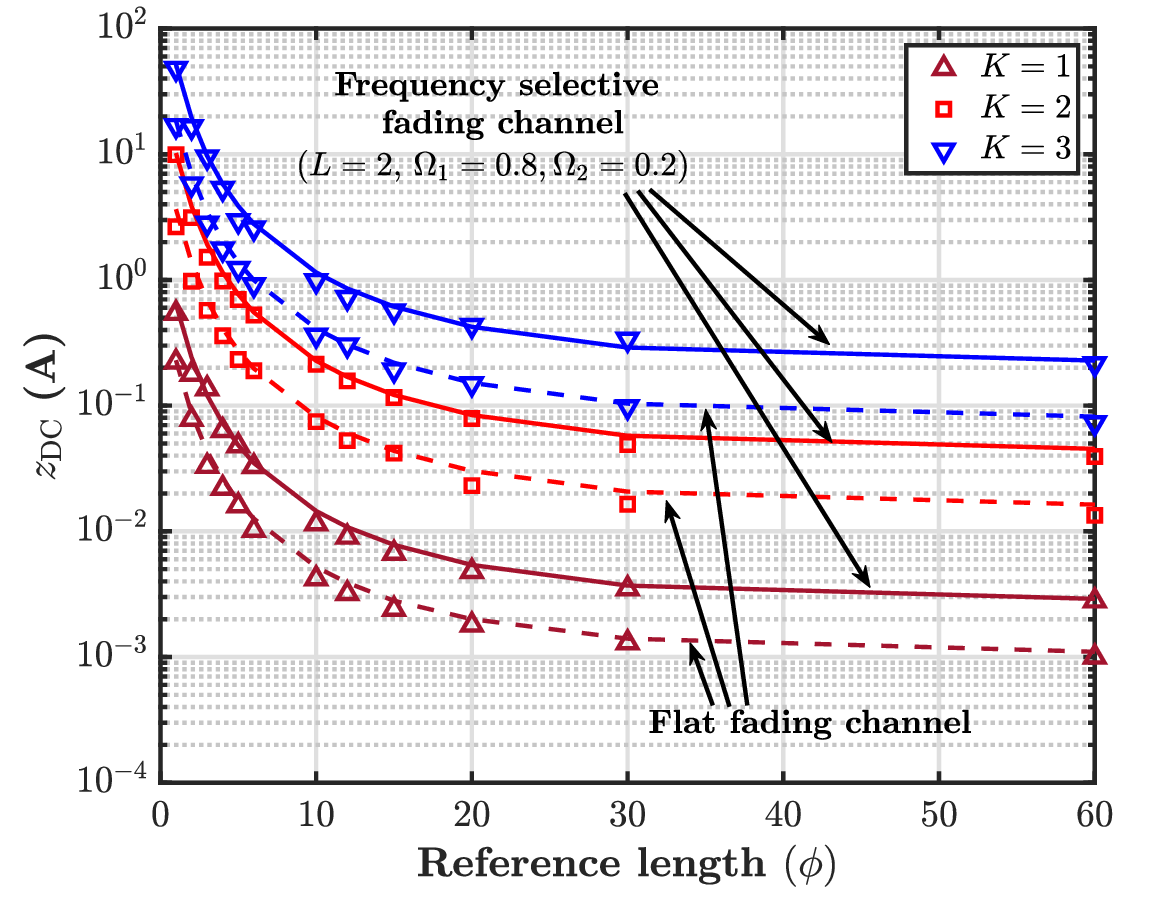}
\vspace{-4mm}
\caption{Impact of $K$ on $z_{\rm DC}$; $\beta=60$ and $m=4$; lines and markers correspond to analytical and simulation results, respectively.}
\vspace{-2mm}
\label{fig:harv}
\end{figure}

Fig. \ref{fig:harv} depicts the EH performance of the transmitted SR-DCSK waveform, where the simulation results (markers) match with the theoretical results (lines). Specifically, the figure demonstrates the variation of $z_{\rm DC}$ with respect to $\phi$, for $K=1,2,$ and $3$, respectively. Both flat and frequency selective fading scenarios have been considered here. We observe that, as claimed in Theorem \ref{theoeh1}, even increasing $K$ by one results in significant enhancement of $z_{\rm DC}$. Moreover, an enhanced reference length has a negative impact on the harvested DC, which corroborates the claim made in \cite{jstsp}. Furthermore, as claimed in \cite{dcskvtc}, here also we observe that the frequency selective nature of the channel is beneficial for WPT when compared against its flat fading counterpart. Finally, when $\phi \in [1,\phi_{\rm opt}]$, Fig. \ref{fig:optphi} and Fig. \ref{fig:harv} provide us an interesting summary, i.e., a trade-off. It illustrates that increasing $\phi$ is beneficial for wireless information transfer (WIT) but detrimental for WPT, and vice-versa. Hence, it is not possible to design a signal that is simultaneously WIT and WPT optimal.

\begin{figure}[!t]\centering\includegraphics[width=\linewidth]{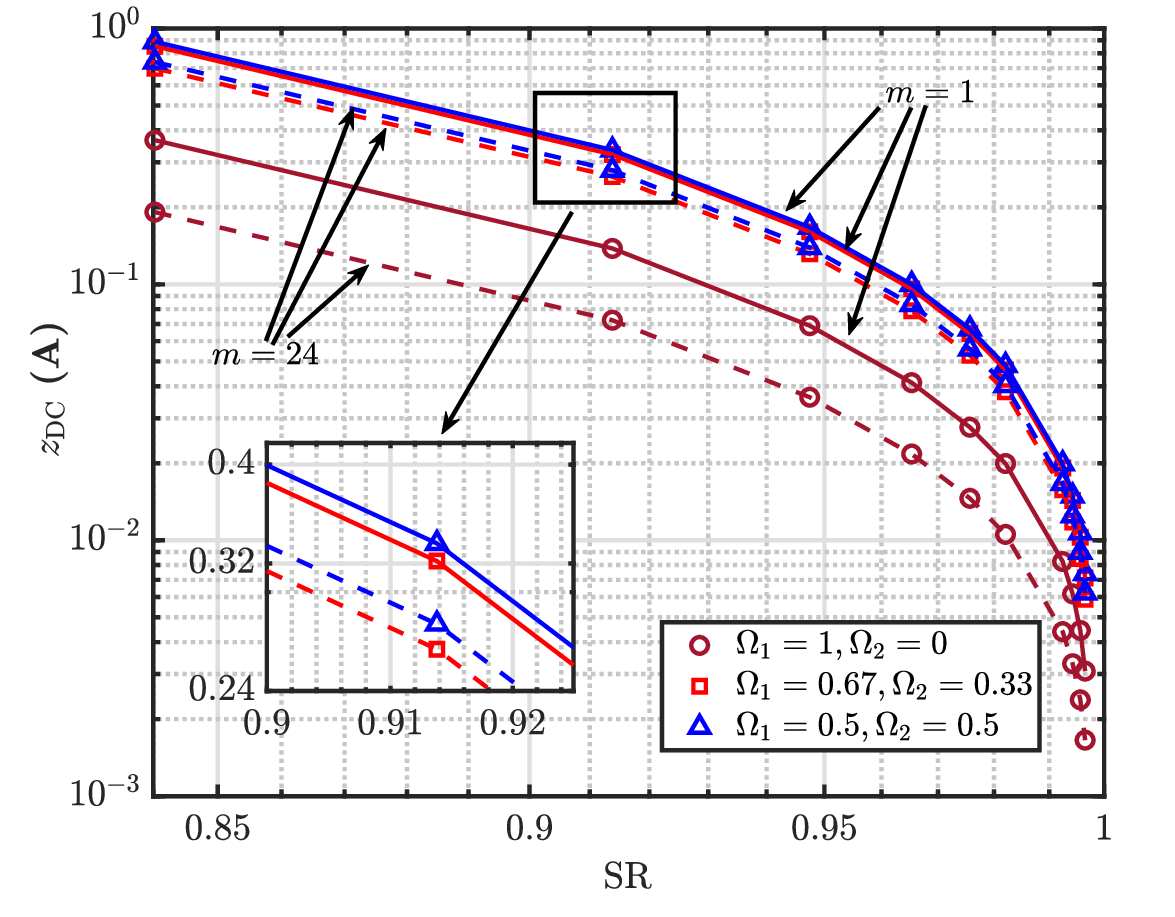}
\vspace{-4mm}
\caption{Effect of channel on ${\rm SR}-z_{\rm DC}$ region; $N=3,M=2,K=1,\gamma_0=12$ dB, and $\beta=60$.}
\vspace{-2mm}
\label{fig:pcdzdc_ch}
\end{figure}

\begin{figure}[!t]\centering\includegraphics[width=\linewidth]{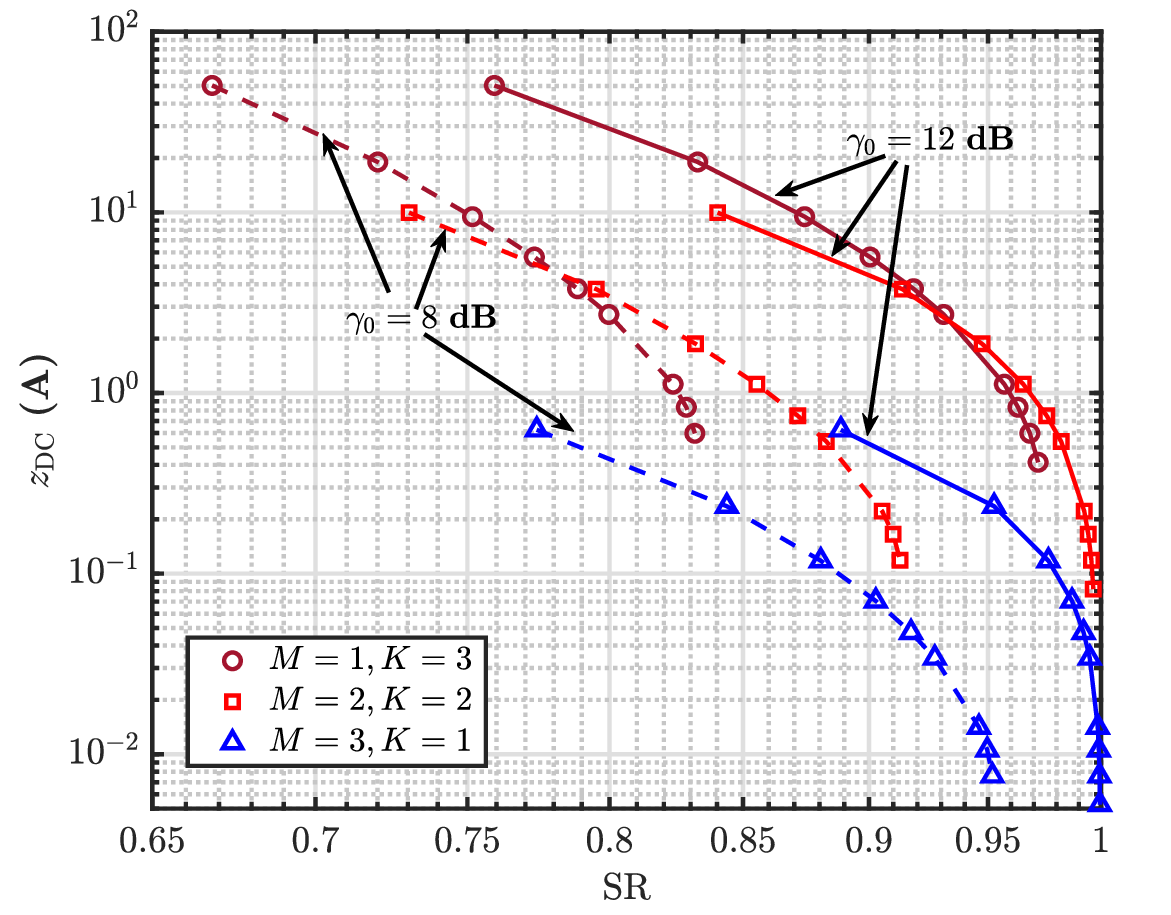}
\vspace{-4mm}
\caption{Effect of antenna choice on ${\rm SR}-z_{\rm DC}$ region; $N=4,m=6,\Omega_1=0.8,\Omega_2=0.2,$ and $\beta=60$.}
\vspace{-2mm}
\label{fig:pcdzdc_ant}
\end{figure}

Fig. \ref{fig:pcdzdc_ch} shows the ${\rm SR}-z_{\rm DC}$ region for the considered SWIPT system with $N=3,M=2,$ and $K=1$ under two cases: fading parameter $m=1$ and $m=24$. The figure considers $\gamma_0=12$ dB and $\beta=60$, which results in an optimal reference length $\phi_{\rm opt}=23.91$ (Theorem \ref{theoit2}) and as $\zeta=\frac{\beta}{\phi} \in \mathbb{Z}^+$, we have $\phi_{\rm opt}=20$. Based on the discussion in Section \ref{wit}, we know that ${\rm SR}$ is monotonically increasing when $\phi \in [1,\phi_{\rm opt}]$ and thus, we obtain the ${\rm SR}-z_{\rm DC}$ region in this figure by varying $\phi$ in this range. We observe that for a given $m$, increasing $\phi$ result in degrading the EH performance, while enhancing the BER response of the system. Moreover, the `degree of channel frequency-selectivity' enhances the ${\rm SR}-z_{\rm DC}$ region irrespective of the value of $m$; for example, decreasing $\Omega_1$ implies that the channel is transitioning from the `flat' to the `frequency-selective' domain, which enhances the region by a significant amount. Indeed this is true, which has also been discussed earlier in the EH performance demonstrated in Fig. \ref{fig:harv}. Furthermore, the figure also depicts that the ${\rm SR}-z_{\rm DC}$ region shrinks with increasing $m$. This observation is inline with the observation made in Section \ref{wpt}, where we note that the best EH performance is achieved with $m=1$ and it monotonically decreases with increasing $m$.

The impact of various $M,K$ combinations (for a fixed $N$) on the proposed ${\rm SR}-z_{\rm DC}$ region is illustrated in Fig. \ref{fig:pcdzdc_ant}, where we specifically consider two cases of $\gamma_0=8$ dB and $\gamma_0=12$ dB. Accordingly, from Theorem \ref{theoit2}, we have $\phi_{\rm opt}=15,20$ corresponding to $\gamma_0=8,12$ dB, respectively. Hence, the ${\rm SR}-z_{\rm DC}$ region for $\gamma_0=8$ dB is obtained with $\phi \in [1,15]$ and we have $\phi \in [1,20]$ for $\gamma_0=12$ dB. Clearly, for a given $\gamma_0$, an increase in $M$ $(K)$ results in enhanced ${\rm SR}$ $(z_{\rm DC})$, and vice-versa. This observation is in line with the claims made in Theorem \ref{theoit2} and Theorem \ref{theoeh1}, where we note that ${\rm SR}$ improves monotonically with increasing $M$ and $z_{\rm DC}$ increases proportionally with $K$ in the order of $K^4$, respectively. Moreover, the figure also demonstrates that for any particular combination of $M$ and $K$, a higher $\gamma_0$ results in greater ${\rm SR}$, and vice-versa.

\begin{figure}
\centering\includegraphics[width=\linewidth]{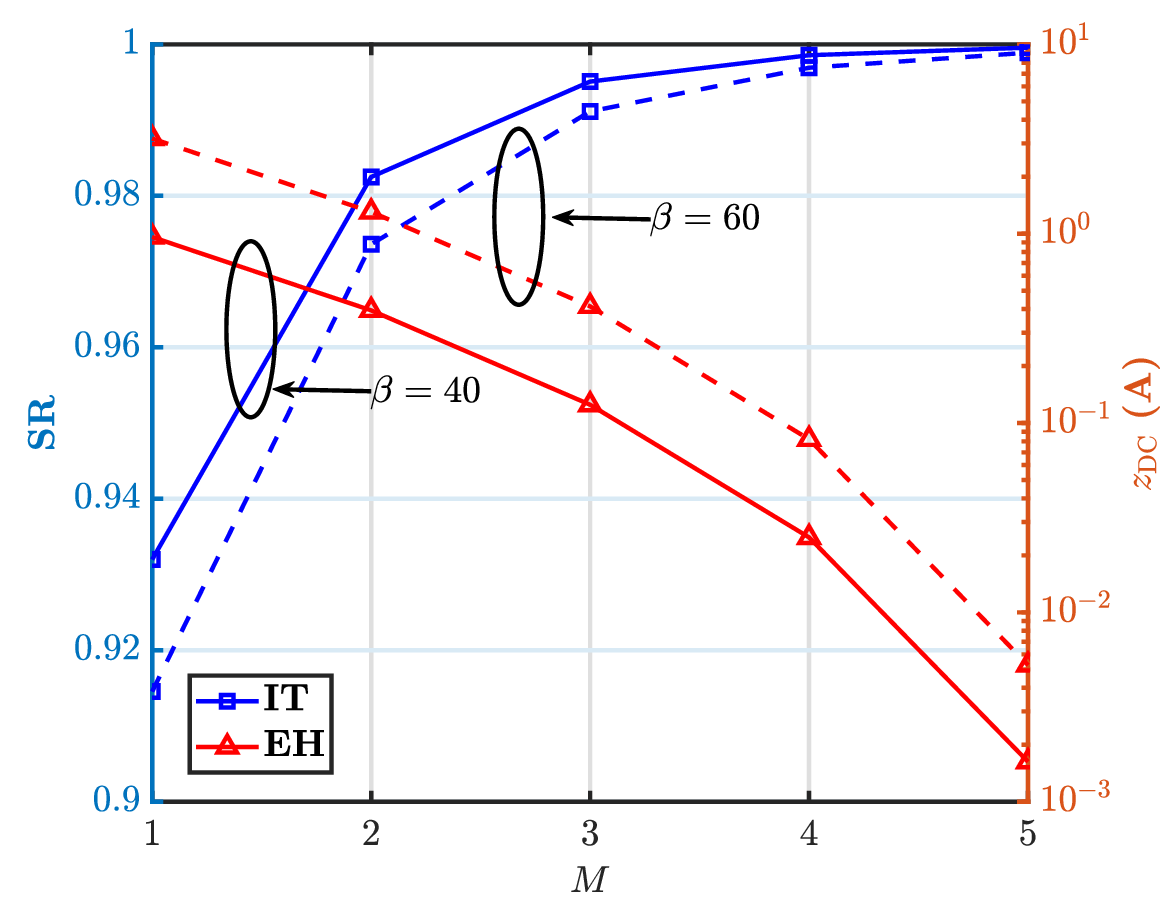}
\vspace{-4mm}
\caption{Impact of antenna choice on SWIPT; $\gamma_0=10$ dB, $m=6,\Omega_1=0.8,\Omega_2=0.2,N=6$ and $K=N-M.$ }
\label{fig:bersr}
\vspace{-2mm}
\end{figure}

Fig. \ref{fig:bersr} illustrates the importance of $M$ and $K$ on the IT and EH performance of the proposed SWIPT architecture. The figure shows that, as claimed in Theorem \ref{theoit2} and Theorem \ref{theoeh1}, the appropriate receiver configuration (i.e., value of $M$ and $K$) is chosen depending on the application specific requirements of EH and/or IT. Moreover, here we also observe that apart from $M$ and $K$, both ${\rm SR}$ and $z_{\rm DC}$ depend on the spreading factor $\beta$ as well. This can be claimed from the fact that even for identical values of $M$ and $K$, both ${\rm SR}$ and $z_{\rm DC}$ result in non-identical values for different $\beta$. Finally, it is interesting to note that the IT performance corresponding to $\beta=40$ is relatively better with respect to its $\beta=60$ counterpart. The reason behind this non-intuitive observation is as follows. From Theorem \ref{theoit2}, we obtain the SR optimal reference length $\phi_{\rm opt}=15.6155$ and $20$ for $\beta=40$ and $60$, respectively. Based on the discussion in Section III, we always choose an appropriate $\phi$ closest to $\phi_{\rm opt}$ as the optimal $\phi$, such that $\frac{\beta}{\phi} \in \mathbb{Z}^+$. Hence, for both $\beta=40$ and $60$, we have optimal $\phi=20$, which eventually leads to higher SR even for a smaller value of $\beta$. Therefore, we can rightly state that an appropriate $\beta$ as well as a right choice of $M$ and $K$ is crucial to  guarantee an enhanced SWIPT performance.

\begin{figure}[!t]\centering\includegraphics[width=\linewidth]{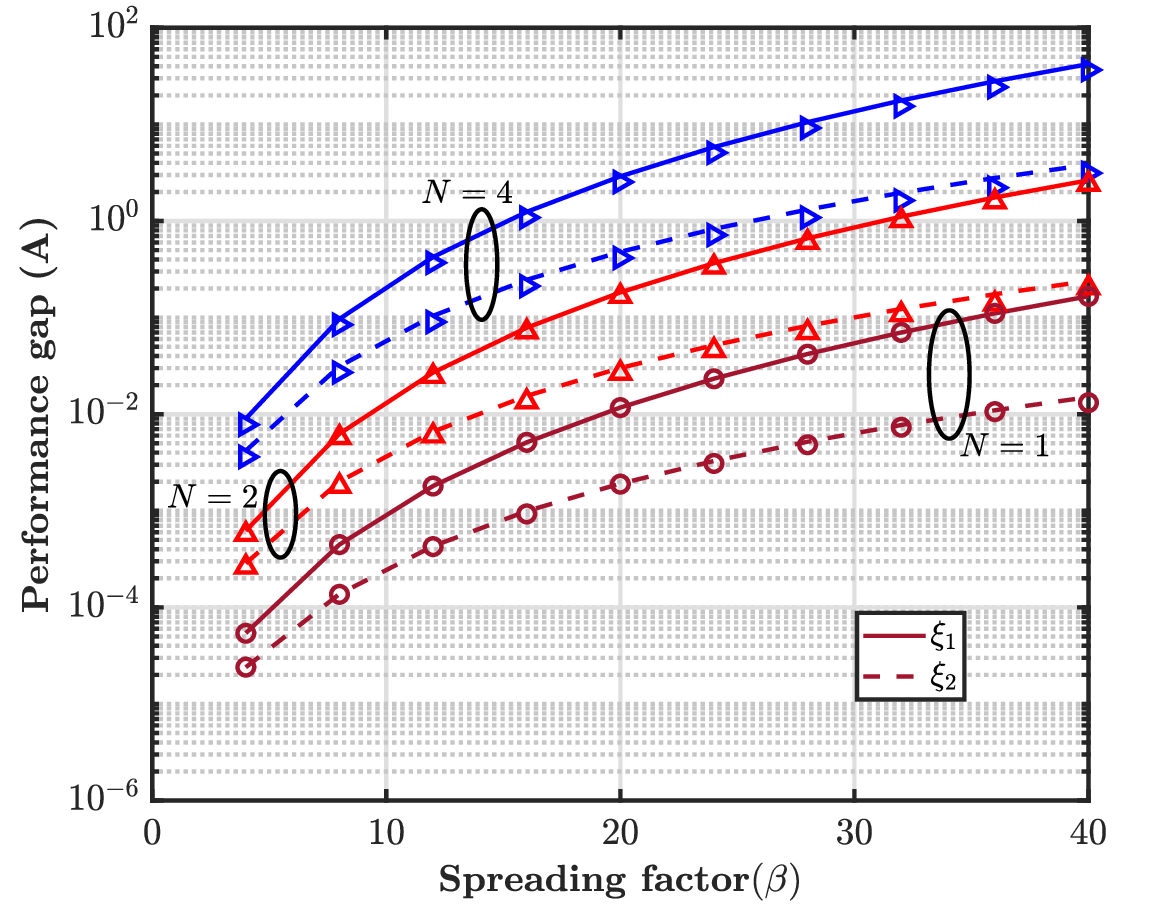}
\vspace{-4mm}
\caption{Effect of information on WPT; $m=4,\Omega_1=0.6,$ and $\Omega_2=0.4$. Lines and markers correspond to analytical and simulation results, respectively.}
\vspace{-2mm}
\label{fig:gap}
\end{figure}

Fig. \ref{fig:gap} depicts the importance of correlation-based waveform design in context of the proposed receiver structure, where all the receiver antennas are being utilized in the EH mode. Specifically, we investigate the performance gap between the waveforms $z_{\rm DC}^{\rm UM},z_{\rm DC}^{\rm opt},$ and $z_{\rm DC}^{\rm PT}$, as stated in Table \ref{tab:summ}, for $N=1,2,$ and $3$, where we have $\xi_1=z_{\rm DC}^{\rm PT}-z_{\rm DC}^{\rm UM}$ and $\xi_2=z_{\rm DC}^{\rm PT}-z_{\rm DC}^{\rm opt}$. The figure demonstrates that to obtain an enhanced EH performance, identical transmit waveform designs cannot be used corresponding to the $K<N$ and $K=N$ scenarios, respectively. In other words, waveforms designed specifically for SWIPT can not be WPT optimal in nature, and vice-versa. Observe that, for identical system parameters and fading scenario, the simulation results (markers) match with their analytical counterparts (lines); this verifies our analysis to characterize $\xi_1$ and $\xi_2$ as obtained in \eqref{pgap}. The figure illustrates that the parameter $\xi_i$ $i \in \{1,2\}$ increases with $\beta$ and moreover, for a given $\beta$, a higher number of receiver antenna results in greater $\xi$. Finally, this observation justifies the importance of proposing the waveform design \eqref{symwpt}, when $K=N$, i.e., the receiver is solely concerned with WPT.

\begin{figure}[!t]\centering\includegraphics[width=\linewidth]{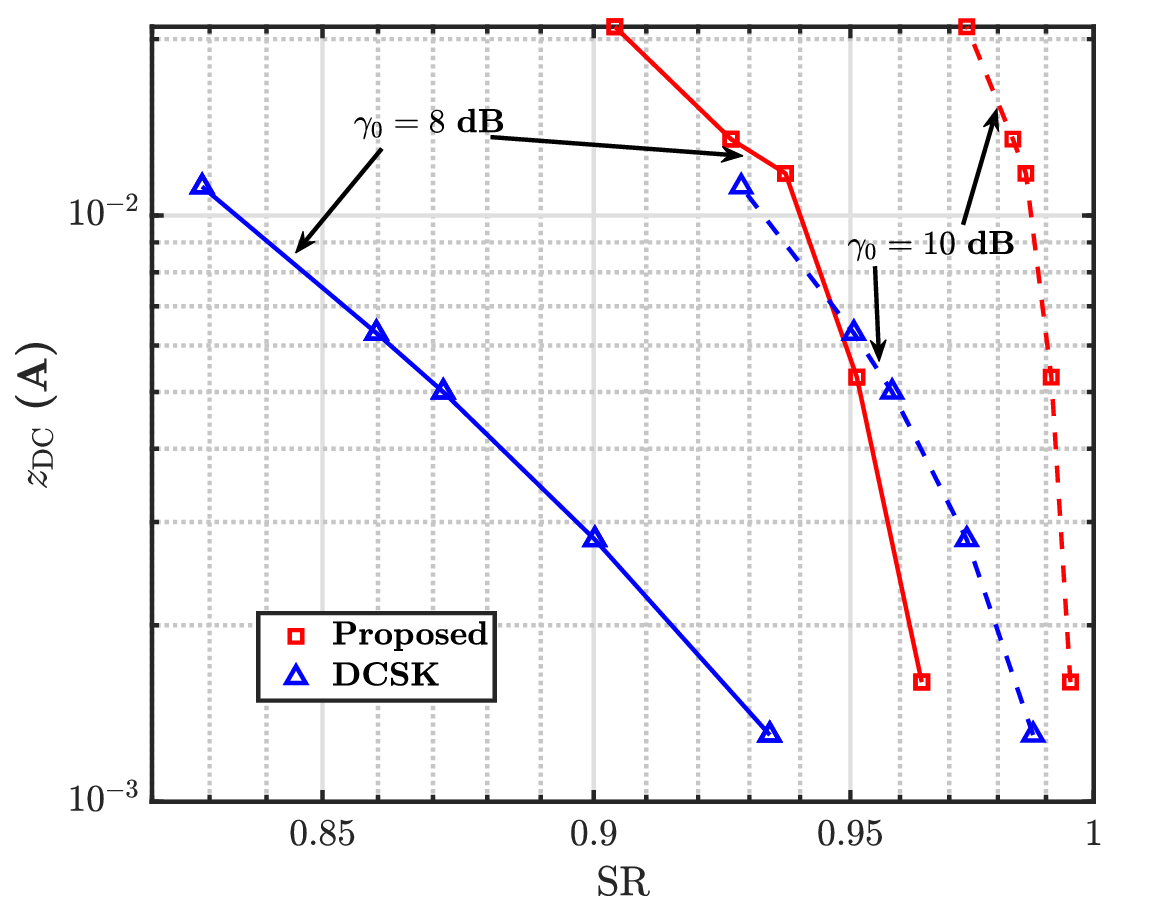}
\vspace{-4mm}
\caption{Impact of the proposed and conventional DCSK-based waveforms on the SR-$z_{\rm DC}$ region; $N=4,M=3,K=1,m=6,\Omega_1=0.8,$ and $\Omega_2=0.2$.}
\vspace{-2mm}
\label{fig:dcomp}
\end{figure}
Fig. \ref{fig:dcomp} compares the ${\rm SR}-z_{\rm DC}$ region of the proposed waveform with that of the conventional DCSK, where we consider $\beta=40,60,80,90,$ and $120$. Note that for DCSK, we have the reference length $\beta$ unlike the proposed waveform, where we obtain the maximum SR for the reference length $\phi$ (Theorem \ref{theoit2}). Accordingly, for each $\beta$, we evaluate the corresponding BER and the harvested DC in this figure, which reveals that, for identical set of system parameters, the proposed waveform results in enhanced SR-$z_{\rm DC}$ region when compared to the conventional DCSK. The reason for this observation is attributed to the intelligent choice of the reference length, which can effectively altered, depending on the application specific requirements of acceptable BER or harvested DC or both.

\section{Conclusion}

In this paper, we investigated the effects of conventional communication-based chaotic waveforms in SWIPT. We considered a SIMO set-up with a single antenna transmitter and a multi-antenna receiver, where the transmitter employs a DCSK-based signal generator. Specifically, depending on the requirement, each receiver antenna can be utilized in either of the IT or EH modes. By taking into account a generalized frequency selective fading and the nonlinearities of the EH process, we characterized the proposed architecture in terms of the BER and harvested DC. We showed that both these metrics are dependent on the parameters of the transmitted waveform and also on the number of the receiver antennas being utilized in the IT and EH mode, respectively. Moreover, we also investigated the BER-energy trade-off to propose different waveform designs corresponding to SWIPT and sole WPT and information transfer, respectively. Numerical results show that the proposed architecture is effective in combining the benefits of chaotic waveform-based signal design and SWIPT. An immediate extension of this work is to investigate the  proposed architecture performance in terms of transmit signal design, where we aim to improve on the data rate, but without compromising on the BER and WPT performance. Moreover, other scenarios can also be considered including one with relays that harvest power and then transmit.

\appendices

\section{Proof of Theorem \ref{theoit1}}\label{app1}
Based on the receiver design, the decision variable is $\delta_l=\sum\limits_{n=1}^{M}\delta_{l,n}$, where $\delta_{l,n}$ is obtained from \eqref{deltadef}. Accordingly, the mean and variance of $\delta_l$ are calculated as follows.
\vspace{-1mm}
\begin{align}  \label{mean}
\mathbb{E}\{\delta_l\}&=\mathbb{E} \left\lbrace \sum\limits_{n=1}^{M} \left( \zeta\phi P_{\rm t}r^{-a}\sum\limits_{i=1}^L \alpha_{i,n}^2 x_{l,q}^2d_l  \right.\right. \nonumber \\
& \!\!\!\!\!\!\!\!\!\!\!\!\!\!\!\!\!\!\left.\left. +\zeta\sum\limits_{q=0}^{\phi-1}w_{q,n} \left( \sqrt{P_{\rm t}r^{-a}}\sum\limits_{i=1}^L \alpha_{i,n} x_{l,q}d_l \right)+ \sum\limits_{p=1}^{\zeta}\sum\limits_{q=0}^{\phi-1} \Bigg( w_{p,q+\phi,n} \right. \right. \nonumber \\
& \!\!\!\!\!\!\!\!\!\!\!\!\!\!\!\!\!\!\left.\left.\left. \times \left( \sqrt{P_{\rm t}r^{-a}}\sum\limits_{i=1}^L \alpha_{i,n} x_{l,q} \right) + w_{p,q+\phi,n}w_{q,n} \right) \right)  \right\rbrace \nonumber \\
&\!\!\!\!\!\!\!\!\!\!\!\!\!\!\!\!\!\!\overset{(a)}{=}\zeta\phi P_{\rm t}r^{-a}d_l \mathbb{E}\{x_k^2\} \sum\limits_{n=1}^{M}\sum\limits_{i=1}^L \alpha_{i,n}^2  \overset{(b)}{=} \frac{\beta r^{-a}d_l\varepsilon_b}{\left( \phi+\beta \right)} \sum\limits_{n=1}^{M}\sum\limits_{i=1}^L \alpha_{i,n}^2,
\end{align}
where $(a)$ follows from the fact that except the first term, all the others have zero mean. Moreover, $(b)$ follows from $\beta=\zeta\phi$ and by using \eqref{bnrgy}. Furthermore, based on the independence property of random variables, we obtain
\vspace{-1mm}
\begin{align}    \label{var}
{\rm Var}\{\delta_l\}&={\rm Var}\left\lbrace \zeta\sum\limits_{n=1}^{M} \sum\limits_{q=0}^{\phi-1}w_{q,n} \left( \sqrt{P_{\rm t}r^{-a}}\sum\limits_{i=1}^L \alpha_{i,n} x_{l,q}d_l \right) \right\rbrace \nonumber \\
&  \!\!\!\!\!\!\!\!\!\!\!\!\!\!\!\!\!\!+ {\rm Var}\left\lbrace \sum\limits_{n=1}^{M}\sum\limits_{p=1}^{\zeta}\sum\limits_{q=0}^{\phi-1} \left( w_{p,q+\phi,n} \left( \sqrt{P_{\rm t}r^{-a}}\sum\limits_{i=1}^L \alpha_{i,n} x_{l,q} \right) \right.\right. \nonumber \\
&\left. \!\!\!\!\!\!\!\!\!\!\!\!\!\!\!\!\!\!+ w_{p,q+\phi,n}w_{q,n} \Bigg) \right\rbrace \nonumber \\
&\!\!\!\!\!\!\!\!\!\!\!\!\!\!\!\!\!\!=\phi\zeta^2 P_{\rm t}r^{-a}\frac{N_0}{2}\mathbb{E}\{x_q^2\} \sum\limits_{n=1}^{M}\sum\limits_{i=1}^L \alpha_{i,n}^2 +\phi\zeta M \frac{N_0^2}{4} \nonumber \\
& \!\!\!\!\!\!\!\!\!\!\!\!\!\!\!\!\!\! + \phi\zeta P_{\rm t}r^{-a}\frac{N_0}{2}\mathbb{E}\{x_q^2\} \sum\limits_{n=1}^{M}\sum\limits_{i=1}^L \alpha_{i,n}^2 \nonumber \\
&\!\!\!\!\!\!\!\!\!\!\!\!\!\!\!\!\!\!= \frac{\beta N_0}{2} \left(  \frac{MN_0}{2} + \frac{r^{-a}\varepsilon_b (\zeta+1)}{\phi+\beta}\sum\limits_{n=1}^{M}\sum\limits_{i=1}^L \alpha_{i,n}^2 \right).
\end{align}
Since both $\delta_l|(d_l=+1)$ and $\delta_l|(d_l=-1)$ are sum of a large number of random variables, it is appropriate to assume them as Gaussian distributed \cite{bookcitingexample}. Furthermore, by assuming equally likely transmission of $d_l=\pm 1$,  the conditional BER (based on channel conditions) is
\vspace{-1mm}
\begin{align} \label{berdef}
{\rm BER}&=\frac{1}{2}\mathbb{P} \left\lbrace \delta_l<1 | d_l=+1 \right\rbrace  + \frac{1}{2}\mathbb{P} \left\lbrace \delta_l>1 | d_l=-1 \right\rbrace \nonumber \\
& =\frac{1}{2}{\rm erfc} \left( \left[\frac{2{\rm Var}\{\delta_l| d_l=+1\}}{\mathbb{E}^2\{\delta_l| d_l=+1\}} \right]^{-\frac{1}{2}}   \right).
\end{align}
Hence, by using  \eqref{mean} and \eqref{var} in \eqref{berdef}, followed by trivial algebraic manipulations, we obtain
\vspace{-1mm}
\begin{equation}  \label{berfg}
{\rm BER}(\kappa)\!=\! \frac{1}{2}{\rm erfc} \left(\! \left[ \frac{M\left(\phi+\beta\right)^2}{2\beta\gamma_0^2\kappa^2}\!+\!\left(\frac{\zeta+1}{\zeta}\right)\!\!\left(\frac{\phi+\beta}{\phi\gamma_0\kappa}\right) \right]^{-\frac{1}{2}}\! \right),
\end{equation}
where we define $\kappa=\sum\limits_{n=1}^{M}\sum\limits_{i=1}^L \alpha_{i,n}^2$.
Hence, the exact BER of the system is
\vspace{-1mm}
\begin{equation}
{\rm BER}=\int_0^{\infty} {\rm BER}(\kappa)f\left( \kappa \right){\rm d}\kappa,
\end{equation}
with $f\left( \kappa \right)$ being the probability distribution function of $\kappa$, which is evaluated as follows.

We define $\kappa=\sum\limits_{n=1}^{M}\sum\limits_{i=1}^L \alpha_{i,n}^2$, where $\alpha_{i,n}$ is a Nakagami-$m$ distributed random variable with parameter $m$ and power gain $\Omega_{i,n}$ such that $\sum\limits_{i=1}^L \Omega_{i,n}=1$ $\forall$ $z=1,\dots,M$. Moreover, we consider identical channel statistics across all the $M$ antennas, i.e., $\Omega_{i,1}=\Omega_{i,2},\dots,\Omega_{i,M}=\Omega_i$ $\forall$ $i=1,\dots,L$. In this context, we define $\kappa_{n}=\sum\limits_{i=1}^L \alpha_{i,n}^2$ $\forall$ $n=1,\dots,M$ and it is well known that the square of a Nakagami-$m$ random variable follows a Gamma distribution \cite{papoulis}. Hence, we have $\alpha_{i,n}^2 \sim \Gamma \left( m, \frac{\Omega_i}{m} \right)$ where a random variable $X$ following a Gamma distribution with parameters $m$ and $\mathbb{E}[X]=\Omega$ is denoted as
\vspace{-1mm}
\begin{equation}
X \sim \Gamma\left( m,\frac{\Omega}{m} \right) \:\:\:\: \text{and} \:\: f(x)=\frac{x^{m-1}{\rm e}^{-\left(\frac{x}{\Omega/m}\right)}}{\Gamma(m)\left( \Omega/m \right)^m}, \quad x \geq 0.
\end{equation}
Accordingly, we obtain $\kappa_n=\sum\limits_{i=1}^L\alpha_{i,n}^2 \sim \Gamma \left( mL, \frac{1}{mL}\right)$. Furthermore, since $\kappa_1,\dots,\kappa_{M}$ are independent, the probability distribution of $\kappa=\sum\limits_{n=1}^{M}\kappa_{n}$ is obtained as $\Gamma \left( MmL, \frac{1}{mL} \right)$, i.e., we have \eqref{gpdf}.

\section{Proof of Theorem \ref{theoit2}}\label{app2}
The AWGN assumption simplifies $\kappa$ defined in \eqref{berf} to $\kappa=M$, which leads to
\vspace{-1mm}
\begin{align}
{\rm BER}&= \frac{1}{2}{\rm erfc}\!\! \left(\! \sqrt{M}\left[ \frac{\left(\phi+\beta\right)^2}{2\beta\gamma_0^2}+\left(\frac{\zeta+1}{\zeta}\right)\left(\frac{\phi+\beta}{\phi \gamma_0}\right) \right]^{-\frac{1}{2}}\! \right) \nonumber \\
&=\frac{1}{2}{\rm erfc} \left(\sqrt{\frac{M}{\Lambda(\phi)}} \right),
\end{align}
where
\vspace{-1mm}
\begin{align}  \label{ldef}
\Lambda(\phi)&=\frac{\left(\phi+\beta\right)^2}{2\beta\gamma_0^2}+\left(\frac{\zeta+1}{\zeta}\right)\left(\frac{\phi+\beta}{\phi \gamma_0}\right)\nonumber \\
&\overset{(a)}{=}\frac{\left(\phi+\beta\right)^2}{\beta\gamma_0}\left(\frac{1}{2\gamma_0}+\frac{1}{\phi}\right).
\end{align}
Here $(a)$ follows from $\beta=\zeta\phi$ and some trivial algebraic manipulations. With  $M$ antennas being utilized in the IT mode and all the other parameters remaining constant, our objective is to minimize the BER and obtain the corresponding reference length, for which the minima is attained. In this context, the first derivative of $\Lambda(\phi)$ with respect to $\phi$ results in
\vspace{-1mm}
\begin{equation}
\Lambda'(\phi)=\frac{\partial\Lambda(\phi)}{\partial \phi}=\frac{\left(\phi+\beta\right)}{\beta\gamma_0} \left( \frac{\phi+2\gamma_0}{\gamma_0\phi}-\frac{\left(\phi+\beta\right)}{\phi^2} \right).
\end{equation}
Moreover, by putting $\Lambda'(\phi)=0$ and considering practical assumptions such as $\gamma_0 \in \mathbb{R}^+$ and $\phi>0$, we obtain $\phi_{\rm opt}=\frac{\gamma_0}{2} \left( \sqrt{1+\frac{4\beta}{\gamma_0}}-1 \right).$ Now we evaluate $\Lambda''(\phi)=\frac{\partial^2\Lambda(\phi)}{\partial\phi^2}$ at $\phi=\phi_{\rm opt}$ to decide on $\phi_{\rm opt}$ being a maxima/minima. It turns out that we have $\Lambda''(\phi_{\rm opt})>0$, which implies $\phi=\phi_{\rm opt}$ to be a minima and accordingly, the minimum value of $\Lambda(\phi)$ is obtained by replacing $\phi=\phi_{\rm opt}$ in \eqref{ldef}. This is equivalent to $\sqrt{\frac{M}{\Lambda(\phi)}}$ having a maxima at $\phi=\phi_{\rm opt}$ and as ${\rm erfc}(x)$ is monotonically decreasing for $x \geq 0$, we can state that the minimum BER obtained is ${\rm BER_{opt}}=\frac{1}{2}{\rm erfc} \left(\sqrt{\frac{M}{\Lambda(\phi_{\rm opt})}} \right).$

\section{Proof of Theorem \ref{theoeh1}}  \label{app3}
From \eqref{zdef1}, the output DC is obtained as
\vspace{-1mm}
\begin{align}  \label{ehtheo1}
z_{\rm DC}&\overset{(a)}{=}k_2R_{ant}P_{\rm t}r^{-a}\mathbb{E}\left\lbrace \left(  \sum\limits_{n=1}^{K}\sum\limits_{k=1}^{\phi+\beta} \sum\limits_{i=1}^L \alpha_{i,n} s_{l,k} \right) ^2 \right\rbrace \nonumber \\
&\qquad+ k_4R_{ant}^2P^2_{\rm t}r^{-2a}\mathbb{E}\left\lbrace \left(  \sum\limits_{n=1}^{K}\sum\limits_{k=1}^{\phi+\beta} \sum\limits_{i=1}^L \alpha_{i,n} s_{l,k} \right) ^4 \right\rbrace \nonumber \\
&\overset{(b)}{=}\nu_1\mathbb{E}\left\lbrace \left(  \sum\limits_{n=1}^{K} \sum\limits_{i=1}^L \alpha_{i,n} \right) ^2 \right\rbrace\mathbb{E}\left\lbrace \left( \sum\limits_{k=1}^{\phi+\beta} s_{l,k} \right) ^2 \right\rbrace \nonumber \\
& \qquad+ \nu_2\mathbb{E}\left\lbrace \left(  \sum\limits_{n=1}^{K} \sum\limits_{i=1}^L \alpha_{i,n} \right) ^4 \right\rbrace\mathbb{E}\left\lbrace  \left( \sum\limits_{k=1}^{\phi+\beta} s_{l,k} \right) ^4\right\rbrace \nonumber \\
&\overset{(c)}{=}\nu_1K^2\mathbb{E}\left\lbrace \left(  \sum\limits_{i=1}^L \alpha_{i} \right) ^2 \right\rbrace\mathbb{E}\left\lbrace \left( \sum\limits_{k=1}^{\phi+\beta} s_{l,k} \right) ^2 \right\rbrace \nonumber \\
&\qquad+ \nu_2K^4\mathbb{E}\left\lbrace \!\! \left(   \sum\limits_{i=1}^L \alpha_{i} \right) ^4 \right\rbrace\mathbb{E}\left\lbrace \!\! \left( \sum\limits_{k=1}^{\phi+\beta} s_{l,k} \right) ^4 \right\rbrace,
\end{align}
where $(a)$ follows from \eqref{approxs}, $(b)$ follows from defining the constants $\nu_1 = r^{-a}k_2R_{ant}P_{\rm t}$ and $\nu_2=r^{-2a}k_4R_{ant}^2P_{\rm t}^2$, and $(c)$ follows from the assumption of identical channel statistics across all the $K$ antennas, i.e., $\Omega_{i,1}=\Omega_{i,2},\dots,\Omega_{i,K}=\Omega_i$ $\forall$ $i=1,\dots,L$. Furthermore, we define $\mathcal{X}=\sum\limits_{k=1}^{\phi+\beta} s_{l,k}$ and $\mathcal{H}=\sum\limits_{i=1}^L \alpha_{i}$ to obtain
\vspace{-1mm}
\begin{equation}
z_{\rm DC}=\nu_1K^2\mathbb{E}\left\lbrace \mathcal{H}^2 \right\rbrace\mathbb{E}\left\lbrace \mathcal{X}^2 \right\rbrace + \nu_2K^4\mathbb{E}\left\lbrace \mathcal{H}^4 \right\rbrace\mathbb{E}\left\lbrace \mathcal{X}^4 \right\rbrace.
\end{equation}
To facilitate this proof, we now evaluate the second and fourth order (raw) moments of $\mathcal{X}$ and $\mathcal{H}$, respectively.

Based on the chaotic frame construction in \eqref{symsr}, we have $\mathcal{X}=\sum\limits_{k=1}^{\phi+\beta} s_{l,k}=\left( 1+\zeta d_l \right)\sum\limits_{k=1}^{\phi}x_k$, where the invariant probability density function (PDF) of $x_k$, is \cite{bookcitingexample}
\vspace{-1mm}
\begin{align}  \label{spdf}
f_X(x)=\begin{cases} 
\dfrac{1}{\pi\sqrt{1-x^2}}, & |x|< 1,\\
0, & \text{otherwise}.
\end{cases}&
\end{align}
Hence, we obtain
\vspace{-1mm}
\begin{align}  \label{s2}
&\mathbb{E}\left\lbrace \mathcal{X} ^2 \right\rbrace=\mathbb{E}\left\lbrace \left( 1+\zeta d_l \right)^2 \right\rbrace\mathbb{E}\left\lbrace \left( \sum\limits_{k=1}^{\phi}x_k \right) ^2 \right\rbrace \nonumber \\
&\overset{(a)}{=}\frac{1}{2}\left( \left(  1+\zeta \right)^2+\left(  1-\zeta \right) ^2 \right)\mathbb{E}\left\lbrace \sum\limits_{k=1}^{\phi}x_k^2 + 2 \!\! \sum_{\substack{k_1,k_2=1 \\ k_1\neq k_2}}^{\phi}x_{k_1}x_{k_2} \right\rbrace \nonumber \\
& =\left(  1+\zeta^2 \right) \!\! \sum\limits_{k=1}^{\phi}\mathbb{E}\left\lbrace x_k^2 \right\rbrace =\frac{\phi}{2}\left(  1+\zeta^2 \right),
\end{align}
where (a) follows from the assumption of equally likely transmission of $d_l=\pm 1$ and the multinomial theorem, and the final result follows from $\mathbb{E}\left\lbrace x_k^2 \right\rbrace=\int_{-1}^{1}\frac{x_k^2dx}{\pi\sqrt{1-x_k^2}}=\frac{1}{2}$ and by using \cite[Eq.~3.67]{bookcitingexample}. Moreover, a similar multinomial theorem-based framework results in
\vspace{-1mm}
\begin{align}  \label{s4}
&\mathbb{E}\left\lbrace \mathcal{X} ^4 \right\rbrace= \mathbb{E}\left\lbrace \left( 1+\zeta d_l \right)^4 \right\rbrace\mathbb{E}\left\lbrace \left( \sum\limits_{k=1}^{\phi}x_k \right) ^4 \right\rbrace\nonumber \\
&\overset{(b)}{=}\left( 1+6\zeta^2+\zeta^4 \right)\!\mathbb{E}\!\left\lbrace \sum\limits_{k_1+k_2+\dots+k_{\phi}=4}\! \frac{4!}{k_1! \: k_2! \: \dots \: k_{\phi}!}\! \prod\limits_{j=1}^{\phi} x_j^{k_j}\!\! \right\rbrace \nonumber \\
&\overset{(c)}{=}\left( 1+6\zeta^2+\zeta^4 \right) \left( \sum\limits_{k=1}^{\phi}\mathbb{E}\left\lbrace x_k^4 \right\rbrace+ 3\mathbb{E}^2\left\lbrace x_k^2 \right\rbrace\phi(\phi-1) \right) \nonumber \\
& \overset{(d)}{=}\frac{3}{8} \left( 1+6\zeta^2+\zeta^4 \right)\left( 2\phi^2-\phi \right),
\end{align}
where $(b)$ follows from the multinomial theorem and $(c)$ and $(d)$  follows from $\mathbb{E}\left\lbrace x_k \right\rbrace=\int_{-1}^{1}\frac{x_kdx}{\pi\sqrt{1-x_k^2}}=0$ and $\mathbb{E}\left\lbrace x_k^4 \right\rbrace=\int_{-1}^{1}\frac{x_k^4dx}{\pi\sqrt{1-x_k^2}}=\frac{3}{8}$, respectively.

\noindent Next, based on the channel model as described in Section \ref{chmodel}, we obtain
\vspace{-1mm}
\begin{align}  \label{a2}
\mathbb{E}\left\lbrace \mathcal{H}^2 \right\rbrace&=\mathbb{E}\left\lbrace  \sum\limits_{i=1}^L \alpha_{i}^2 + 2 \sum_{\mathclap{\substack{i_1,i_2=1\\i_1 \neq i_2}}}^L \alpha_{i_1}\alpha_{i_2}\right\rbrace \nonumber \\
&=1+\frac{2}{m}\left(\frac{\Gamma(m+0.5)}{\Gamma(m)}\right)^2 \sum_{\mathclap{\substack{i_1,i_2=1\\i_1 \neq i_2}}}^L \sqrt{\Omega_{i_1}\Omega_{i_2}}.
\end{align}
Here we use the $n$-th order moment of $\alpha_i$, i.e.,
\vspace{-1mm}
\begin{align}    \label{val1}
\mathbb{E}\{\alpha_i^n\}&=\frac{2m^m}{\Gamma(m)\Omega_i^m}\int_0^{\infty} z^{2m+n-1}e^{-\frac{mz^2}{\Omega_i}} dz \nonumber \\
&\!\!\!\!\!\!\!\!\!\!\!\!\!\!\!\!\!\!\!\!=\frac{1}{\Gamma(m)} \left(\frac{\Omega_i}{m}\right)^{\frac{n}{2}}\!\! \int_0^{\infty}\!\!\!\! \omega^{m+\frac{n}{2}-1}e^{-\omega}d\omega =\frac{\Gamma(m+\frac{n}{2})}{\Gamma(m)}\left(\frac{\Omega_i}{m}\right)^{\frac{n}{2}},
\end{align}
which follows from the transformation $ \frac{mz^2}{\Omega_i} \rightarrow \omega$.
Based on \eqref{val1} and the multinomial theorem as demonstrated in \eqref{s4}, we obtain
\vspace{-1mm}
\begin{align}  \label{a4}
\mathbb{E}\left\lbrace \mathcal{H}^4 \right\rbrace&=\sum\limits_{i_1+i_2+\dots+i_L=4} \frac{4!}{i_1! \: i_2! \: \dots \: i_L!} \nonumber \\
& \times \prod\limits_{j=1}^L  \frac{\Gamma(m+\frac{i_j}{2})}{\Gamma(m)}\left(\frac{\Omega_j}{m}\right)^{\frac{i_j}{2}}.
\end{align}
Finally, by replacing \eqref{s2}, \eqref{s4}, \eqref{a2}, and \eqref{a4} in \eqref{ehtheo1}, we get \eqref{harvnp}.

\section{Proof of Proposition \ref{theoeh2}}\label{app5}
From \eqref{ehtheo1}, we have
\vspace{-1mm}
\begin{equation}
z_{\rm DC}=\nu_1K^2\mathbb{E}\left\lbrace \mathcal{H}^2 \right\rbrace\mathbb{E}\left\lbrace \mathcal{X}^2 \right\rbrace + \nu_2K^4\mathbb{E}\left\lbrace \mathcal{H}^4 \right\rbrace\mathbb{E}\left\lbrace \mathcal{X}^4 \right\rbrace.
\end{equation}
The flat fading channel is a special case of its frequency selective counterpart with $\alpha_1$ following unit power Nakagami-$m$ distribution, i.e., $\Omega_1=1$ and $\alpha_i=0$ $\forall$ $i>1$. As a result, we obtain
\vspace{-1mm}
\begin{equation}
z_{\rm DC}=\nu_1K^2\mathbb{E}\left\lbrace \alpha_1^2 \right\rbrace \mathbb{E}\left\lbrace \mathcal{X} ^2 \right\rbrace + \nu_2K^4\mathbb{E}\left\lbrace \alpha_1^4 \right\rbrace\mathbb{E}\left\lbrace \mathcal{X}^4 \right\rbrace.
\end{equation}
The channel coefficient $\alpha_1$ being unit power implies $\mathbb{E}\left\lbrace \alpha_1^2 \right\rbrace=1$ and by using $i=1,n=4,$ and $\Omega_1=1$ in \eqref{val1}, we have $\mathbb{E}\left\lbrace \alpha_1^4 \right\rbrace=\frac{\Gamma(m+2)}{\Gamma(m)}\left(\frac{1}{m}\right)^{2}=\frac{m+1}{m}$. Hence, by combining these along with \eqref{s2} and \eqref{s4}, we get $z_{\rm DC, FF}$ in \eqref{flat}. Furthermore, a no fading scenario implies $m \rightarrow \infty$, i.e., we obtain $z_{\rm DC, NF}=\lim\limits_{m \rightarrow \infty}z_{\rm DC, FF}$ as stated in \eqref{nof}.

\bibliographystyle{IEEEtran}
\bibliography{refs}

\end{document}